\documentstyle[prd,aps,eqsecnum,preprint,tighten]{revtex}
\input epsf

\begin{document}

\preprint{\vbox to 30 pt{
\hbox{CPT-2000/PE.4053}
\hbox{gr-qc/0009034}
\vfil}}

\title{Scalar-tensor gravity in an accelerating universe}

\author{G.~Esposito-Far\`ese$^{a,b}$ and D.~Polarski$^{c,b,d}$}

\address{\ }
\address{$^a$Centre de Physique Th\'eorique, CNRS Luminy,
Case 907, F 13288 Marseille cedex 9, France}
\address{$^b$D\'epartement d'Astrophysique Relativiste
et de Cosmologie,\\
Observatoire de Paris-Meudon, F 92195 Meudon cedex, France}
\address{$^c$Lab.~de Math\'ematique et Physique Th\'eorique,
UPRES-A 6083 CNRS,\\
Universit\'e de Tours, Parc de Grandmont, F 37200 Tours, France}
\address{$^d$Laboratoire de Physique Math\'ematique et Th\'eorique,
UMR 5825 CNRS\\ Universit\'e de Montpellier II, F 34095 Montpellier
cedex 05, France}
\date{September 11, 2000}
\maketitle

\begin{abstract}
We consider scalar-tensor theories of gravity in an accelerating
universe. The equations for the background evolution and the
perturbations are given in full generality for any parametrization of
the Lagrangian, and we stress that apparent singularities are sometimes
artifacts of a pathological choice of variables. Adopting a
phenomenological viewpoint, {\it i.e.}, from the observations back to
the theory, we show that the knowledge of the luminosity distance as a
function of redshift up to $z \sim (1-2)$, which is expected in the
near future, severely constrains the viable subclasses of scalar-tensor
theories. This is due to the requirement of positive energy for both
the graviton and the scalar partner. Assuming a particular form for the
Hubble diagram, consistent with present experimental data, we
reconstruct the microscopic Lagrangian for various scalar-tensor
models, and find that the most natural ones are obtained if the
universe is (marginally) closed.
\end{abstract}


\draft

\section{Introduction}
Recently, there has been a lot of interest in cosmological solutions
in the presence of a cosmological constant, when the latter is
significant compared to the present total energy density of the
universe. Indeed, the Hubble diagram based on observations of type Ia
supernovae up to a redshift $z\sim 1$ seems to imply that our
universe is presently accelerating \cite{Perl,Garn}. These data,
when combined with the observed location of the first acoustic peak
of the CMB temperature fluctuations, favor a spatially flat universe
whose energy density is dominated by a ``cosmological
constant''-like term. The flatness of the Universe is corroborated
by the latest Boomerang and Maxima data \cite{Bo,Ma}, in accordance
with the inflationary paradigm, though a
marginally closed Universe is still allowed by
the position of the first acoustic (Doppler) peak at $l\sim 200$.
A significant cosmological constant may help in
resolving the dark matter problem --~for dustlike matter alone
observations seem to imply $\Omega_m\sim~0.3$~-- and in reconciling
flat Cold Dark Matter (CDM) models with observations in the framework
of $\Lambda$CDM models. Finally, a cosmological constant is an
elegant way to allow a high Hubble constant $H_0$ with
$h \equiv H_0/(100\ {\rm km}\ {\rm s}^{-1}\ {\rm Mpc}^{-1})\approx
0.65$ and a sufficiently old universe $t_0>11$Gyr \cite{BOPS}
(see also, e.g., \cite{SS} for a recent comprehensive review and
references therein).

Therefore, this interpretation, if confirmed by future observations,
constitutes a fundamental progress towards the solution of the dark
matter problem and the formation of large-scale structure in the
Universe out of primordial fluctuations generated by some inflationary
model. That is certainly what makes it so appealing and gives it, maybe
somehow prematurely, the status of new paradigm. A striking consequence
for our Universe is then its present acceleration, for a large range of
equations of state \cite{CP00}.

Of course, from the point of view of particle physics, a pure
cosmological constant of the order of magnitude $\Lambda
\approx 3\times 10^{-122}\, c^3/(\hbar G)$, interpreted as
the vacuum energy, is extremely problematic. This is why attempts
were made to find some alternative explanation to the origin of the
acceleration under the form of some scalar field $\Phi$ (sometimes
called quintessence \cite{CDS98}, ``$\Lambda$''-field, etc.) whose
slowly varying energy density would mimic an effective cosmological
constant. This is very reminiscent of the mechanism producing the
inflationary phase itself with the fundamental difference that this
scalar field, which does not have to be {\it a priori\/} the
inflaton, is accelerating the expansion today, therefore at a much
lower energy scale. This of course has problems of its own as this
effective cosmological constant term started dominating the universe
expansion only in the very recent past (the so-called ``cosmic
coincidence'' problem). Indeed, the energy density of the field
$\Phi$ must remain subdominant at very early stages and come to
dominate in the recent past only. Hence, specific evolution
properties are required to meet these constraints and were indeed
shown to hold for particular potentials, partly alleviating the
problem of the initial conditions. For inverse power-law potentials
the energy density of the scalar field was shown to decrease less
rapidly than the background energy density so that it can be
negligible in the early universe and still come to dominate in the
recent past \cite{RaPe88}. For exponential potentials,
\cite{FeJo98,RaPe88} the scalar field energy density has the very
interesting behavior that it tends to a fixed fraction of the total
energy density, these are the so-called ``tracker solutions''. Hence
a pure exponential potential is excluded if data confirm that the
energy density of the scalar field is dominating today, as this
fraction had to be small at the time of nucleosynthesis. A slightly
different potential is proposed in \cite{ZWS99} and a classification
of the scaling behavior of the scalar field for various potentials
has been given in \cite{LiSc98}. Hence, though a minimally coupled
scalar field is an attractive possibility, some degree of fine tuning
still remains in the parameters of the potential
\cite{LiSc98,KoLy98}.

If one admits that it is some minimally coupled scalar field which
plays the role of an effective cosmological constant while gravity is
described by general relativity, the question immediately arises: What
is the ``right'' potential $U(\Phi)$ of this scalar field? In a recent
work by Starobinsky \cite{St98}, the following ``phenomenological''
point of view was adopted: Instead of looking for more or less well
motivated models, like the interesting possibilities discussed above,
it is perhaps more desirable to extract as much information as possible
from the observations (a similar approach can also be adopted to
reconstruct the inflaton potential) in order to reconstruct the
scalar field potential, if the latter exists at all. Cosmological
observations could then be used to constrain the particle physics
model in which this scalar field is supposed to originate. In the
context of general relativity plus a minimally coupled scalar field,
it was shown that the reconstruction of $U(\Phi)$ can be implemented
once the quantity $D_L(z)$, the luminosity distance as a function of
redshift, is extracted from the observations \cite{St98,HT99},
something that is expected in the near future.\footnote{Actually, it
is shown in Ref.~\cite{SRSS} that the potential $U(\Phi)$ can already
be reconstructed from present experimental data, although not yet
very accurately.} The SNAP (Supernovae Acceleration Probe) satellite
will notably make measurements with an accuracy at the percent level
up to $z\approx 1.7$. Of course, in this way only the recent past of
our Universe, up to redshifts $z\sim (1-2)$ (for reference, we will
push some of our simulations up to $z\sim 5$), is probed and so the
reconstruction is made only for the corresponding part of the
potential. Crucial information is therefore gained on the microscopic
Lagrangian of the theory through relatively ``low'' redshift
cosmological observations.

A further step is to generalize the same mechanism in the framework
of scalar-tensor theories of gravity, sometimes called ``generalized
quintessence''. The usual minimally coupled models are certainly
ruled out if, for example, it turns out that this component of the
energy density obeys an equation of state $p=w\rho$ with $w<-1$
($\rho\ge 0$). Strangely enough, such an unexpected equation of state
which in itself implies new physics, is in fair agreement with the
observations \cite{Cald}. Also the inequality $dH^2(z)/dz \ge 3
\Omega_{m,0} H_0^2 (1+z)^2$ must hold for a minimally coupled scalar
field, hence its violation would force us to consider more complicated
theories, possibly scalar-tensor theories. There are also strong
theoretical motivations. These theories, in which the scalar field
participates in the gravitational interaction, are the most natural
alternatives to general relativity (GR). Indeed, scalar partners to
the graviton generically arise in theoretical attempts at quantizing
gravity or at unifying it with other interactions. For instance, in
superstrings theory, a dilaton is already present in the
supermultiplet of the 10-dimensional graviton, and several other
scalar fields (called the moduli) also appear when performing a
Kaluza-Klein dimensional reduction to our usual spacetime. Moreover,
contrary to other alternative theories of gravity, scalar-tensor
theories respect most of GR's symmetries: conservation laws,
constancy of (non-gravitational) constants, local Lorentz invariance
(even if a subsystem is influenced by external masses), and they
also have the capability of satisfying the weak equivalence
principle (universality of free fall of laboratory-size objects)
even for a strictly massless scalar field. Nevertheless, they can
describe many possible deviations from GR, and their
predictions have been thoroughly studied in various situations:
solar-system experiments \cite{willbook,def1,def4}, binary-pulsar
tests \cite{willbook,def1,def7}, gravitational-wave detection
\cite{will94,def9}. Finally these scalar-tensor theories could play
a crucial role in the very early universe, for example in the Pre
Big Bang inflationary model (see e.g. \cite{GaVe93}).

Thus, in this work we are investigating the possibility to have an
accelerating universe in the context of scalar-tensor theories of
gravity instead of pure GR. This has indeed attracted a lot of
interest recently and such cosmological models have been studied and
possibly confronted with observations like CMB anisotropies, or
the growth of energy density perturbations (see for instance
\cite{dn93,w98,dp99,nsa99,uz99,ch99,ck99,pbm99,hw99,bp99,GaLo00}).
However, we emphasize once more that the central point of view
adopted here, in analogy with Starobinsky \cite{St98}, is to
constrain the model with the {\it experimental} knowledge of the
Hubble diagram up to $z\sim (1-2)$. This is precisely why use of the
redshift $z$ as basic variable is crucial for our purpose: Quantities
like $H(z)$ are directly observable, in contrast to,
say,\footnote{The function $H(t)$ can be obtained from the knowledge
of $H(z)$ thanks to the relation $t=-\int dz/[(1+z)H(z)]$, but the
directly observable quantity is $H(z)$.}
$H(t)$ or $H(\Phi)$. For instance, we have access to $H(z)$ through
the direct measurement of the luminosity distance in function of
redshift $D_L(z)$. In a recent letter \cite{beps00}, it was shown
that the knowledge of both $H(z)$ and $\delta_m(z)$ is sufficient to
reconstruct the full theory (again, in the range probed by the data).
This means that we do {\it not} choose any specific theory {\it a
priori}, but instead we reconstruct whatever theory possibly realized
in Nature.

As we will see, the knowledge of $H(z)$ on its own, though
insufficient in order to fully reconstruct a scalar-tensor theory
unless one makes additional assumptions, turns out to be already very
constraining when subclasses of models are considered. This is
particularly interesting because it means that cosmological
observations at low redshifts implying an accelerated expansion might
well give new constraints on scalar-tensor theories. We will show
that this is indeed the case.

Throughout the paper, we use natural units for which $\hbar=c=1$, and
the signature $\scriptstyle (-+++)$, together with the sign
conventions of \cite{mtw73}. In Section II, we introduce the
general formalism of scalar-tensor theories of gravity and their
different parametrizations. In Section III, we briefly review the
severe experimental restrictions imposed on these theories {\it
today}. In Section IV, we consider FRW universes in the framework of
scalar-tensor gravity and we give the equations for the different
parameterizations. In Section V, we review the full reconstruction
problem. In Section VI, we give a detailed study of subclasses of
models, which are investigated using the background equations.
Finally, in Section VII, our results are summarized and discussed.

\section{Scalar-tensor theories of gravity}
We are interested in a universe where gravity is described by a
scalar-tensor theory, and we consider the action \cite{bnw70}
\begin{equation}
S={1\over 16\pi G_*} \int d^4x \sqrt{-g}
\Bigl(F(\Phi)~R -
Z(\Phi)~g^{\mu\nu}
\partial_{\mu}\Phi
\partial_{\nu}\Phi
- 2U(\Phi) \Bigr)
+ S_m[\psi_m; g_{\mu\nu}]\ .
\label{S_JF}
\end{equation}
Here, $G_*$ denotes the bare gravitational coupling constant (which
differs from the measured one, see Eq.~(\ref{Geff}) below), $R$ is
the scalar curvature of $g_{\mu\nu}$, and $g$ its determinant. In
Ref.~\cite{beps00}, we used different conventions (corresponding to
the choice $8\pi G_* = 1$ in the above action); here, the quantity
$F(\Phi)$ is dimensionless. This factor $F(\Phi)$ needs to be positive
for the gravitons to carry positive energy. The action of matter
$S_m$ is a functional of some matter fields $\psi_m$ and of the metric
$g_{\mu\nu}$, but it does not involve the scalar field $\Phi$. This
ensures that the weak equivalence principle is exactly satisfied.

The dynamics of the real scalar field $\Phi$ depends {\it a
priori\/} on three functions: $F(\Phi)$, $Z(\Phi)$, and the potential
$U(\Phi)$. However, one can always simplify $Z(\Phi)$ by a
redefinition of the scalar field, so that $F(\Phi)$ and $Z(\Phi)$ can
be reduced to only one unknown function. Two natural parametrizations
are used in the literature: (i)~the Brans-Dicke one, corresponding to
$F(\Phi) = \Phi$ and $Z(\Phi) = \omega(\Phi)/\Phi$; and (ii)~the
simple choice $Z(\Phi) = 1$ and $F(\Phi)$ arbitrary. This second
parametrization is however sometimes pathological. [The derivatives of
$\Phi$ can become imaginary in perfectly regular situations; see
the discussion about Eq.~(\ref{positiveEnergy}) below.] In the
following, we will write the field equations in terms of the two
functions $F(\Phi)$ and $Z(\Phi)$, so that any particular choice can
be recovered easily.

The variation of action (\ref{S_JF}) gives straightforwardly
\begin{mathletters}
\begin{eqnarray}
F(\Phi) \left(R_{\mu\nu}-{1\over2}g_{\mu\nu}R\right)
&=& 8\pi G_* T_{\mu\nu}
+ Z(\Phi) \left(\partial_\mu\Phi\partial_\nu\Phi
- {1\over 2}g_{\mu\nu}
(\partial_\alpha\Phi)^2\right)
\nonumber\\
&&+\nabla_\mu\partial_\nu F(\Phi) - g_{\mu\nu}\Box F(\Phi)
- g_{\mu\nu} U(\Phi)\ ,
\label{einstein}\\
2Z(\Phi)~\Box\Phi &=&
-{dF\over d\Phi}\,R - {dZ\over d\Phi}\,(\partial_\alpha\Phi)^2
+ 2 {dU\over d\Phi}\ ,
\label{BoxPhi}\\
\nabla_\mu T^\mu_\nu &=& 0\ ,
\label{matter}
\end{eqnarray}
\label{2.2}
\end{mathletters}
where $T \equiv T^\mu_\mu$ is the trace of the
matter energy-momentum tensor $T^{\mu\nu} \equiv (2/\sqrt{-g})\times
\delta S_m/\delta g_{\mu\nu}$. The scalar-field equation
(\ref{BoxPhi}) can of course be rewritten differently if one uses
the trace of Eq.~(\ref{einstein}) to replace the curvature scalar
$R$ by its source, and one gets the Brans-Dicke-like equation
\begin{equation}
2\varpi~\Box\Phi =
8\pi G_* {dF\over d\Phi}\,T
-{d\varpi\over d\Phi}\,(\partial_\alpha\Phi)^2
- 4 U\,{dF\over d\Phi} + 2 {dU\over d\Phi}\,F\ ,
\label{oldBoxPhi}
\end{equation}
where $2\varpi\equiv 2ZF+3(dF/d\Phi)^2$. [In the Brans-Dicke
representation where $F = \Phi$ and $Z = \omega(\Phi)/\Phi$, this
factor $2\varpi$ reduces to the well-known expression $2\omega(\Phi) +
3$.] In the following, we will however use the form (\ref{BoxPhi}),
which will simplify considerably our calculations.

The above equations are written in the so-called Jordan frame (JF).
Since in action (\ref{S_JF}), matter is universally coupled to
$g_{\mu\nu}$, this ``Jordan metric'' defines the lengths and times
actually measured by laboratory rods and clocks (which are made of
matter). All experimental data will thus have their usual
interpretation in this frame. In particular, the observed Hubble
parameter $H$ and the measured redshifts $z$ of distant objects are
Jordan-frame quantities.

However, it is usually much clearer to analyze the equations and the
mathematical consistency of the solutions in the so-called Einstein
frame (EF), defined by diagonalizing the kinetic terms of the
graviton and the scalar field. This is achieved thanks to a
conformal transformation of the metric and a redefinition of the
scalar field. Let us call $g^*_{\mu\nu}$ and $\varphi$ the new
variables, and define
\begin{mathletters}
\begin{eqnarray}
g^*_{\mu\nu} &\equiv& F(\Phi)~g_{\mu\nu}\ , \label{g*}\\
\left({d\varphi\over d\Phi}\right)^2 &\equiv& {3\over
4}\left({d\ln F(\Phi)\over d\Phi}\right)^2 + {Z(\Phi)\over
2F(\Phi)}\, \label{varphi}\\
A(\varphi) &\equiv& F^{-1/2}(\Phi)\ ,\label{A}\\
2V(\varphi) &\equiv& U(\Phi)~F^{-2}(\Phi)\ .\label{V}
\end{eqnarray}
\label{2.4}
\end{mathletters}
Action (\ref{S_JF}) then takes the form
\begin{equation}
S={1\over 4\pi G_*} \int d^4x \sqrt{-g_*} \left({R^*\over 4} -
{1\over 2} g_*^{\mu\nu} \partial_{\mu}\varphi \partial_{\nu}\varphi
- V(\varphi) \right)
+ S_m[\psi_m; A^2(\varphi)~g^*_{\mu\nu}]\ ,
\label{S_EF}
\end{equation}
where $g_*$ is the determinant of $g^*_{\mu\nu}$, $g_*^{\mu\nu}$
its inverse, and $R^*$ its scalar curvature. Note that the first term
looks like the action of general relativity, but that matter is now
explicitly coupled to the scalar field $\varphi$ through the conformal
factor $A^2(\varphi)$. Quantities referring to the Einstein frame will
always have an asterisk (either in superscript or in subscript), e.g.
$\nabla^*_\mu$ and $\Box^*$ for the covariant derivative and the
d'Alembertian with respect to the Einstein metric. The indices of
Einstein-frame tensors will also be lowered and raised with the
Einstein metric $g^*_{\mu\nu}$ and its inverse $g_*^{\mu\nu}$. The
field equations deriving from action (\ref{S_EF}) take the simple form
\begin{mathletters}
\begin{eqnarray}
R^*_{\mu\nu} - {1\over 2} R^* g^*_{\mu\nu} &=&
8\pi G_* T^*_{\mu\nu}
+ 2 \partial_\mu\varphi\partial_\nu\varphi -
g^*_{\mu\nu}(g_*^{\alpha\beta}
\partial_\alpha\varphi\partial_\beta\varphi)
- 2 V(\varphi) g^*_{\mu\nu}\ ,
\label{einstein*}\\
\Box^*\varphi &=& -4\pi G_*\alpha(\varphi)~T_*
+dV(\varphi)/d\varphi\ ,
\label{Boxvarphi}\\
\nabla^*_\mu T^\mu_{*\nu} &=& \alpha(\varphi)~T_*
\partial_\nu\varphi\ ,
\label{matter*}
\end{eqnarray}
\label{2.6}
\end{mathletters}
where
\begin{equation}
\alpha(\varphi) \equiv {d \ln A \over d\varphi}
\label{alpha}
\end{equation}
is the coupling strength of the scalar field to matter sources
\cite{def1}, and $T_* \equiv g^*_{\mu\nu}T_*^{\mu\nu}$ is the trace
of the matter energy-momentum tensor $T_*^{\mu\nu} \equiv
(2/\sqrt{-g_*})~\delta S_m/\delta g^*_{\mu\nu}$ in Einstein-frame
units. From its definition, one can deduce the relation $T^*_{\mu\nu}
= A^2(\varphi)~T_{\mu\nu}$ with its Jordan-frame counterpart.

Let us underline that the Cauchy problem is well posed in the Einstein
frame \cite{def1}, because all the second-order derivatives of the
fields are separated in the left-hand sides of Eqs.~(\ref{2.6}),
whereas they are mixed in the JF equations (\ref{2.2}). Action
(\ref{S_EF}) also shows that the helicity-2 degree of freedom is
described by the fluctuations of the Einstein metric $g^*_{\mu\nu}$
(whose kinetic term is the standard Einstein-Hilbert one), and that the
EF scalar $\varphi$ is the true helicity-0 degree of freedom of the
theory (since its kinetic term has the standard form). On the other
hand, the fluctuations of the Jordan metric $g_{\mu\nu}$ actually
describe a {\it mixing\/} of helicity-2 and helicity-0 excitations, and
the JF scalar $\Phi$ is related to the helicity-0 degree of freedom via
the complicated relation (\ref{varphi}), because its kinetic term in
action (\ref{S_JF}) comes not only from the naive contribution
$Z(\Phi)\, (\partial_\mu\Phi)^2$ but also from the cross term
$F(\Phi)\,R$. In conclusion, the theory can be mathematically
well defined only if it is possible to write the EF action
(\ref{S_EF}), notably with its negative sign for the scalar-field
kinetic term (so that $\varphi$ carries positive energy). If it happens
that the transformation (\ref{2.4}) is singular for particular values
of $\Phi$, the consistency of the theory should be analyzed in the EF.
Some singularities may be artifacts of the parametrization which is
chosen to write action (\ref{S_JF}), and may not have any physical
significance. On the other hand, Jordan-frame quantities may look
sometimes regular while there is an actual singularity in the Einstein
frame (a typical example is provided when $F(\Phi)$ vanishes). In this
case, the solution should be considered as mathematically inconsistent.
In the following, we will see that the JF is better suited than the EF
for our cosmological study, but we will always check the consistency of
our results by finally translating them in terms of Einstein-frame
quantities.

\section{Known experimental constraints}
The predictions of general relativity in weak-field conditions, and
at present, are confirmed by solar-system experiments at the $0.04
\%$ level \cite{vlbi,dw96}. One should therefore verify that the
scalar-tensor models we are considering are presently close enough
to Einstein's theory.

If the scalar field is very massive (say, if $d^2V/d\varphi^2$ is
large with respect to the inverse of the astronomical unit), its
influence is exponentially small in solar-system experiments, even
if it is strongly coupled to matter. This situation corresponds to
the particular scalar-tensor model considered in Ref.~\cite{fiz99}
(namely $F(\Phi) = \Phi$ and $Z(\Phi)=0$ in action (\ref{S_JF}), but
assuming a large enough value for $d^2U/d\Phi^2$). Although this
situation is phenomenologically acceptable, it remains somewhat
problematic from a field theoretical viewpoint, since the massive
scalar would {\it a priori\/} desintegrate into lighter (matter)
particles.

On the contrary, if the scalar mass is small with respect to the
inverse solar-system distances, it must be presently very weakly
coupled to matter for the theory to be consistent with experimental
data. At the first post-Newtonian order ($1/c^2$ with respect to the
Newtonian interaction), the deviations from general relativity can
be parametrized by two real numbers, that Eddington \cite{edd23}
denoted as $(\beta-1)$ and $(\gamma-1)$. In the present framework,
they take the form \cite{willbook,def1,def4}
\begin{mathletters}
\begin{eqnarray}
\gamma - 1 &=& -2 {\alpha^2\over 1+\alpha^2}= - {(dF/d\Phi)^2\over ZF
+ 2(dF/d\Phi)^2}~,
\label{gammaPPN}\\
\beta - 1 &=& {1\over
2}\,{\alpha^2\over(1+\alpha^2)^2}\,{d\alpha\over d\varphi} = {1\over
4}\,{F~(dF/d\Phi)\over 2 ZF + 3(dF/d\Phi)^2}\, {d\gamma\over
d\Phi}\ ,
\label{betaPPN}
\end{eqnarray}
\label{3.1}
\end{mathletters}
where the first expressions are given in terms of the
Einstein-frame notation (\ref{S_EF})-(\ref{alpha}), whereas the last
ones correspond to the Jordan-frame general representation
(\ref{S_JF}). To simplify, the second expression of
Eq.~(\ref{betaPPN}) has been written in terms of the derivative
of (\ref{gammaPPN}) with respect to $\Phi$.

Using the upper bounds on $(\gamma-1)$ from solar-system
measurements~\cite{vlbi}, we thus get the constraint
\begin{equation}
2\alpha_0^2 \approx (ZF)_0^{-1}~(dF/d\Phi)_0^2 < 4\times 10^{-4}~,
\label{dF0}
\end{equation}
where an index $0$ means the present value of the corresponding
quantity. On the other hand, the experimental bounds on $(\beta-1)$
cannot be used to constrain the derivative $(d\alpha/d\varphi)_0$
appearing in Eq.~(\ref{betaPPN}), since it is multiplied by a factor
$\alpha_0^2$ consistent with 0. Because of nonperturbative
strong-field effects, binary-pulsar tests are however directly
sensitive to this derivative, {\it i.e.}, to the ratio
$-4(\beta-1)/(\gamma-1)$. In a generic class of scalar-tensor models,
Refs.~\cite{def7,def9} have obtained the bound
\begin{equation}
(d\alpha/d\varphi)_0 > -4.5~.
\label{beta0}
\end{equation}

{}From action Eq.~(\ref{S_JF}), one can naively define Newton's
gravitational constant as the inverse factor of the curvature scalar
$R$:
\begin{equation}
G_N \equiv G_* A^2 = G_* / F\ .
\label{G_N}
\end{equation}
However, $G_N$ does not have the same physical meaning as Newton's
gravitational constant in GR. Indeed, the actual Newtonian force
measured (in Cavendish-type experiments) between two close test
masses $m_1$ and $m_2$ is of the form $G_{\rm eff} m_1 m_2/r^2$,
where the effective gravitational constant reads
\cite{willbook,def1,def4}
\begin{equation}
G_{\rm eff} \equiv
G_* A^2 (1+\alpha^2)
= {G_*\over F}
\left({2ZF+4(dF/d\Phi)^2\over2ZF+3(dF/d\Phi)^2}\right)\ .
\label{Geff}
\end{equation}
The contribution $G_* A^2$ is due to the exchange of a graviton
between the two bodies, whereas $G_* A^2 \alpha^2 = G_*
(dA/d\varphi)^2$ comes from the exchange of a scalar particle
between them. Of course, when the distance between the bodies
becomes larger than the inverse mass of the scalar field, its
influence becomes negligible and one gets $G_{\rm eff}\approx G_N$.
Note that as usual, the last expression in Eq.~(\ref{Geff}), in
terms of Jordan-frame notation, is much more complicated than its
Einstein-frame counterpart. In the particular Brans-Dicke
representation, $F = \Phi$ and $Z = \omega(\Phi)/\Phi$, it however
reduces to the simpler (and well-known) form $G_{\rm eff} = G_*
\Phi^{-1}(2\omega+4)/(2\omega+3)$.

The experimental bound (\ref{dF0}) shows that the present
values of $G_{\rm eff}$ and $G_N$ differ by less than $0.02 \%$.
However, they can {\it a priori\/} differ significantly in the past.
It should be noted that the experimental limit on the time
variation of the gravitational constant, $|\dot G_{\rm eff}/G_{\rm
eff}|< 6\times 10^{-12}~{\rm yr}^{-1}$~\cite{dw96}, does {\it not\/}
imply any constraint on $2\dot A/A = -\dot F/F$. Indeed,
$G_{\rm eff}$ can be almost constant even if $A$ (or $F$) varies
significantly. A simple example is provided by Barker's theory
\cite{barker78}, in which $A(\varphi) = \cos\varphi$~: One gets
$G_{\rm eff} = G_* (\cos^2\varphi + \sin^2\varphi) = G_*$, which is
strictly constant independently of the time variations of
$A(\varphi(t))$. Nevertheless, as pointed out in \cite{beps00}, under
reasonable cosmological assumptions, one can derive $G_{\rm
eff}\approx G_N$ with $\sim 10 \%$ accuracy up to redshifts $z\sim
1$.

\section{Scalar-tensor cosmology}
The equations derived in this section generalize those of our
previous paper \cite{beps00} in several ways. First, we use the
most general representation (\ref{S_JF}) of the theory, instead of
the simpler choice $Z=1$ that was made in \cite{beps00}. Second, we
take into account a possible spatial curvature of the universe, which
will be an interesting possibility in our studies of Sec.~VI
below. Third, we write the equations for an arbitrary pressure of
the perfect fluid describing matter in the universe. This will not
be useful for our reconstruction program of the following sections,
as matter can be assumed to be simply dustlike for the redshifts $z
\lesssim 5$ that we will consider, but these general equations may
be interesting for further cosmological studies of earlier epochs of
the universe. Finally, we comment on the Einstein-frame version of
these equations, which are mathematically simpler, but actually more
difficult to use for our purpose.

\subsection{Background}
We consider a Friedmann-Robertson-Walker (FRW) universe whose
background metric in the Jordan frame is given by
\begin{mathletters}
\begin{eqnarray}
ds^2 &=& -dt^2 + a^2(t)~d\ell^2\ ,
\label{ds}\\
d\ell^2 &=& {dr^2\over 1-\kappa r^2}
+ r^2 \left(d\theta^2 + \sin^2\theta~d\phi^2\right)\ ,
\label{dl2}
\end{eqnarray}
\label{4.1}
\end{mathletters}
where $\kappa = -1$, $0$, or $1$ for spatially
open, flat, or closed universes respectively. The scalar field
$\Phi$ (or $\varphi$, in the EF) is also assumed to depend only on
time. Since the relation between the EF and JF is given by $ds^2 =
A^2(\varphi)~ds_*^2$, see Eqs.~(\ref{2.4}), our universe is still of
the FRW type in the EF, with $ds_*^2 = -dt_*^2 +
a_*^2(t_*)~d\ell^2\label{ds*}$ and
\begin{equation}
dt = A(\varphi)~dt_*\ ,
\quad a = A(\varphi)~a_*\ .
\label{aVSa*}
\end{equation}
In the following, matter will be described by a perfect fluid, and
we will write its energy-momentum tensor as
\begin{equation}
T_{\mu\nu} = (\rho+p) u_{\mu} u_{\nu} + p g_{\mu\nu} =
A^{-2}~T^*_{\mu\nu}~= A^{-2}\left((\rho_*+p_*) u^*_{\mu}
u^*_{\nu} + p_* g^*_{\mu\nu}\right)\ ,
\label{T}
\end{equation}
where $u^{\mu} = dx^\mu/|ds|$ and $u_*^{\mu} = dx^\mu/|ds_*|$ are
the spacetime components of the four-dimensional unit velocity of
matter, in JF and EF units respectively. As we are interested in a
FRW background, the spatial components $u_i$ and $u^*_i$ ($i=1, 2,
3$) all vanish. From (\ref{T}), we deduce the relation between the
matter density and pressure in both frames:
\begin{equation}
\rho_* = A^4~\rho\ ,
\quad p_* = A^4~p\ .
\label{rho*VSrho}
\end{equation}

The background equations in the JF follow from
(\ref{einstein})--(\ref{matter}), and read
\begin{mathletters}
\begin{eqnarray}
3F\cdot\left(H^2+{\kappa\over a^2}\right)
&=&
8\pi G_* \rho
+{1\over 2}Z \dot\Phi^2 - 3 H \dot F + U\ ,
\label{H2}\\
-2F\cdot\left(\dot H - {\kappa\over a^2}\right)
&=&
8\pi G_*(\rho+p) + Z \dot\Phi^2 +\ddot F - H\dot F\ ,
\label{dotH}\\
Z\cdot(\ddot\Phi+3H\dot\Phi)
&=&
3{dF\over d\Phi}\left(\dot H + 2 H^2 +{\kappa\over a^2}\right)
- {dZ\over d\Phi}\, {\dot\Phi^2\over 2}
- {dU\over d\Phi}\ ,
\label{ddotPhi}\\
\dot\rho + 3H~(\rho+p) &=& 0\ ,
\label{dmatter}
\end{eqnarray}
\label{4.5}
\end{mathletters}
where $H \equiv d(\ln a)/dt$, and a dot denotes differentiation with
respect to the Jordan-frame time $t$. As usual, if $p/\rho \equiv w =
{\rm const.}$, Eq.~(\ref{dmatter}) is trivially integrated as
$\rho \propto a^{-3(1+w)}$ (and in particular $\rho \propto a^{-3}$
for dustlike matter). Equation~(\ref{ddotPhi}) is actually a
consequence of the other three, and we will not need it in the
following.

Since these equations correspond to the most general parametrization
(\ref{S_JF}) of scalar-tensor theories, many particular cases are
easily recovered. For instance, the case of a minimally coupled
scalar field \cite{St98} is obtained for constant values of $F$ and
$Z$ (say, $F =1$ and $Z = 8\pi G_*$), and the particular model
considered in \cite{fiz99} is recovered immediately for $F = \Phi$
and $Z = 0$.

The corresponding background equations in the EF are very similar to
those in general relativity. They follow from Eq.~(\ref{einstein*}),
and read
\begin{mathletters}
\begin{eqnarray}
3\left(H_*^2 + {\kappa\over a_*^2}\right) &=& 8\pi G_* \rho_*
+ \left({d\varphi\over dt_*}\right)^2
+ 2V(\varphi)\ ,\\
- {3\over a_*}\,{d^2a_*\over dt_*^2}
&=&
4\pi G_*(\rho_*+3p_*) + 2 \left({d\varphi\over dt_*}\right)^2
- 2 V(\varphi)\ ,
\label{acc}
\end{eqnarray}
\label{4.6}
\end{mathletters}
where $H_* \equiv d(\ln a_*)/dt_*$ is the Einstein-frame Hubble
parameter. It is obvious from (\ref{acc}) that a vanishing potential
$V(\varphi)$ implies $d^2a_*/dt_*^2 < 0$, so that the universe is
decelerating in the Einstein frame. However, because of the
relation $a = A(\varphi)~a_*$, see Eq.~(\ref{aVSa*}), the {\it
observed\/} (Jordan-frame) expansion rate $\ddot a$ may be positive
even in this case, and we will see concrete examples in Sec.~VI.A
below. This is an important point to remember: Although we are
looking for cosmological FRW backgrounds whose expansion is
accelerating, the sign of $d^2a_*/dt_*^2$ is {\it a priori\/} not
fixed.

The scalar-field equation of motion in the EF follows from
Eq.~(\ref{Boxvarphi}), and reads
\begin{equation}
\frac{d^2 \varphi}{d t_*^2} + 3 H_* \frac{d\varphi}{d t_*} +
\frac{dV(\varphi)}{d\varphi} = - 4\pi G_*
\alpha(\varphi)~(\rho_*-3p_*)~.
\label{KG}
\end{equation}
It is also similar to the usual Klein-Gordon equation, with the
notable difference of a source term on the right-hand side, with the
coupling strength $\alpha(\varphi)$ defined in Eq.~(\ref{alpha})
above.

It is tempting to tackle our problem in the EF as the equations are
simpler and we can rely on experience gained in general relativity.
However, a crucial difficulty that we encounter is that all physical
quantities which appear in the EF background equations are not those
that come from observations. Moreover, the behavior of matter in the
EF is complicated by the relations (\ref{rho*VSrho}): Instead of the
simple power law $\rho\propto a^{-3}$ for dustlike matter in the JF,
one gets $\rho_* = A^4\rho \propto A a_*^{-3}$ in the EF, where
$A(\varphi(a_*))$ can have {\it a priori\/} any shape. To avoid
these problems, we will thus work in the JF, and show that the
``reconstruction'' program can equally well be implemented, like in
general relativity, although it is mathematically very different. We
will nevertheless check at the end the consistency of the solutions
obtained by translating them in terms of EF quantities.

\subsection{Perturbations}
We now consider the perturbations in the longitudinal gauge.
For this problem, we will restrict our discussion to the case of a
spatially flat FRW universe ($\kappa = 0$), and write the JF
and EF metrics as
\begin{mathletters}
\begin{eqnarray}
ds^2= -(1 + 2 \phi) dt^2 + a^2 (1 - 2\psi) d{\bf x}^2~,
\label{dsPert}\\
ds_*^2= -(1 + 2 \phi_*) dt^2 + a_*^2 (1 - 2\psi_*) d{\bf x}^2~.
\label{ds*Pert}
\end{eqnarray}
\label{4.8}
\end{mathletters}
In the EF, the perturbation equations deriving from
Eq.~(\ref{einstein*}) are strictly the same as in general relativity
plus a minimally coupled scalar field. One thus finds notably
$\phi_* = \psi_*$. On the other hand, the equations for scalar-field
and matter perturbations are modified by the matter-scalar coupling,
proportional to $\alpha(\varphi)$ in Eqs.~(\ref{Boxvarphi}) and
(\ref{matter*}).

For our purpose, it will be more useful to write the perturbation
equations in the (physical) JF. Let us define the gauge invariant
quantity\footnote{Note that our definition differs from the
quantity $\epsilon_m$ introduced in \cite{bardeen80}: $\epsilon_m =
(1+p/\rho)\delta_m$.}
\begin{equation}
\delta_m \equiv {\delta\rho\over \rho +p} + 3Hv\ ,
\label{defdeltam}
\end{equation}
where $v$ is the matter peculiar velocity potential (such that
$\delta u_\mu = -\partial_\mu v$ is the perturbation of the
four-dimensional unit velocity $u_\mu$). We now work in Fourier
space, and assume a spatial dependence
$\exp(i {\bf k}
\cdot {\bf x})$, with
$k
\equiv |{\bf k}|$. The conservation equations of matter
(\ref{matter}) give
\begin{mathletters}
\begin{eqnarray}
\dot\delta_m &=&
-\frac{k^2}{a^2}\, v + 3\,{d(\psi + H v)\over dt}~,\label{conserv_0}\\
\phi &=& \dot v +{p\over \rho}\,(2Hv-\delta_m)~.
\label{conserv_z}
\end{eqnarray}
\label{4.10}
\end{mathletters}
On the other hand, the Einstein equations (\ref{einstein}) give
\begin{mathletters}
\begin{eqnarray}
\psi &=& \phi + \delta F/F\ ,
\label{psi}\\
2F(\dot\psi+H\phi) + \dot F \phi &=& 8\pi G_* (\rho + p) v +
Z \dot\Phi\, \delta\Phi + \dot{\delta F} - H\, \delta F\ ,
\label{dotpsi}\\
-3 \dot F \dot\phi - \left( 2{k^2\over a^2} F - Z \dot\Phi^2 +
3H\dot F \right)\phi &=&
8\pi G_* (\rho+p)\delta_m
+ \left( {k^2\over a^2}-6H^2 -3{\dot F^2\over F^2} \right)\delta F
+ \delta U
\nonumber\\
&&+ Z \dot\Phi \dot{\delta\Phi}
+ 3HZ\dot\Phi\,\delta\Phi
+{1\over 2}\delta Z\,\dot\Phi^2
+ 3{\dot F\over F}\,\dot{\delta F}\ .
\label{fluctmetr}
\end{eqnarray}
\label{4.11}
\end{mathletters}
Note that $\phi\ne \psi$ in the JF, in contrast to the corresponding
problem in general relativity or in the EF. Equation (\ref{psi}) is
actually an obvious consequence of the relation between
$g^*_{\mu\nu}$ and $g_{\mu\nu}$, Eq.~(\ref{g*}), and of the fact that
$\phi_*= \psi_*$. Finally, Eq.~(\ref{BoxPhi})
yields the equation for the dilaton fluctuations $\delta\Phi$:
\begin{eqnarray}
&&\ddot{\delta\Phi} + \left(3H+{d\ln Z\over d\Phi}\,
\dot\Phi\right)\dot{\delta\Phi}\nonumber\\
&&+ \left[{k^2\over a^2} - 3(\dot H + 2H^2){d\over
d\Phi}\left({1\over Z}\,{dF\over d\Phi}\right)
+{d\over d\Phi}\left({1\over Z}\,{dU\over d\Phi}\right)
+{d^2\ln Z\over d\Phi^2}\,{\dot\Phi^2\over 2}
\right]\delta\Phi =
\nonumber\\
&&= \left[{k^2 \over a^2}(\phi-2\psi)
- 3(\ddot\psi+4H\dot\psi+H\dot\phi)\right]{1\over Z}\,{dF\over d\Phi}
+ (3\dot\psi+\dot\phi)\dot\Phi - 2\,{\phi\over Z}\,{dU\over d\Phi}\ .
\label{fluctPhi}
\end{eqnarray}
In the particular representation $Z=1$ used in
Ref.~\cite{beps00}, this equation reduces to the simpler form
\begin{eqnarray}
\ddot{\delta\Phi} + 3H\,\dot{\delta\Phi}
&+& \left[{k^2\over a^2} - 3(\dot H + 2H^2){d^2F\over d\Phi^2}
+{d^2U\over d\Phi^2}\right]\delta\Phi =
\nonumber\\
&=& \left[{k^2 \over a^2}(\phi-2\psi)
- 3(\ddot\psi+4H\dot\psi+H\dot\phi)\right]{dF\over d\Phi}
+ (3\dot\psi+\dot\phi)\dot\Phi - 2\phi\,{dU\over d\Phi}\ .
\label{fluctPhiZ1}
\end{eqnarray}

\section{The reconstruction problem}
The reconstruction of the potential $U(\Phi)$ was shown in
\cite{St98} to be possible in the framework of general relativity
plus a minimally coupled scalar field, the $\Lambda$-field or
quintessence, provided the Hubble diagram (and thus also $H(z)$) can
be extracted from the observations. An essential difference arises
when one deals with scalar-tensor theories: We have to reconstruct
two unknown functions instead of one, hence we need to extract two
quantities (as functions of the redshift $z\equiv a_0/a -1$) from the
observations. Actually, in the minimally coupled case, the knowledge
of the luminosity distance $D_L$ and of the clustering of matter
$\delta_m$, both in function of $z$, provides two independent ways to
reconstruct the scalar field potential \cite{St98}.\footnote{More
precisely, to reconstruct the potential $U(\Phi)$ without any
ambiguity in the minimally coupled case, one needs to know both
$D_L(z)$ and the present energy density of dustlike matter
$\Omega_{m,0}$, or both $\delta_m(z)$ and the present value of the
Hubble constant $H_0$. In our general scalar-tensor case, we need to
know the two functions $D_L(z)$ and $\delta_m(z)$, but no independent
measurement of $\Omega_{m,0}$ or $H_0$ is necessary.} In our case,
both quantities are necessary and the reconstruction itself is
significantly more complicated.

The present section generalizes our previous results of
Ref.~\cite{beps00} not only by considering the most general
parametrization (\ref{S_JF}) of scalar-tensor theories and by taking
into account the possible spatial curvature of the universe, but also
by discussing particular cases that were excluded in this reference.
{}From now on, we will restrict our discussion to the case of a
pressureless perfect fluid ($p=0=p_*$), because all matter in the
universe will be assumed to be simply dustlike, of course besides
that part needed to account for the present accelerated expansion
({\it i.e.}, the scalar field in the present framework).

\subsection{Background}
The first step of the reconstruction program is the same as in
general relativity, since it is purely kinematical and does not
depend on the field content of the theory: If the luminosity distance
$D_L$ is experimentally determined as a function of the redshift
$z$, one can deduce the quantity $H(z)$ from the relation
\begin{equation}
{1\over H(z)} = \left({D_L(z)\over
1+z}\right)'\times\left[1+\Omega_{\kappa,0}\left({H_0
D_L(z)\over 1+z}\right)^2\right]^{-1/2},
\label{HvsDL}
\end{equation}
where the prime denotes the derivative with respect to $z$. The large
square brackets contain a corrective factor involving the
present energy contribution $\Omega_{\kappa,0} \equiv
-\kappa/(a_0^2H_0^2)$ of the spatial curvature of the universe. It
was not written explicitly in Refs.~\cite{SS,St98}, which focused
their discussions on the flat-space case ($\Omega_{\kappa,0} = 0$),
but it is a straightforward consequence of Eqs.~(23)--(25) of
Ref.~\cite{SS}. Since present experimental data suggest that
$|\Omega_{\kappa,0}|$ is small, the flat-space expression for $1/H(z)
= [D_L(z)/(1+z)]'$ is {\it a priori} a good approximation anyway.
Note that even if one uses the exact equation (\ref{HvsDL}), it
reduces to the flat-space expression for $z=0$ (because $D_L(0) =
0$), and therefore $H_0$ is always known without any ambiguity. To
determine $H(z)$ precisely at higher $z$, one then needs to know both
$D_L(z)$ and $\Omega_{\kappa,0}$.

By eliminating $Z\dot\Phi^2$ from the background equations (\ref{H2})
and (\ref{dotH}), we then obtain the equation
\begin{equation}
\ddot F + 5 H \dot F + 2\left(\dot H+ 3 H^2 + {2\kappa\over
a^2}\right)F = 8\pi G_*\rho + 2 U\ ,
\label{ddotF}
\end{equation}
which, when rewritten in terms of the redshift $z$, gives the
fundamental equation
\begin{eqnarray}
F'' + \left[(\ln H)'-\frac{4}{1+z}\right]~F' +
\left[\frac{6}{(1+z)^2} -
\frac{2}{1+z}(\ln H)'-4\left({H_0\over
H}\right)^2\Omega_{\kappa,0}\right]~F =&&\nonumber\\
= \frac{2 U}{(1+z)^2 H^2} + 3 (1+z) \left({H_0\over
H}\right)^2 F_0 \Omega_{m,0}\ .&&
\label{F}
\end{eqnarray}
As before, an index $0$ means the present value of the corresponding
quantity, and we use again the notation $f'\equiv df/dz$. In this
equation, $\Omega_{m,0} \equiv 8\pi G_* \rho_0 / (3 F_0 H_0^2)$ stands
for the present energy density of dustlike matter relative to the
critical density $\varepsilon_{\rm crit}\equiv 3H_0^2/8\pi G_{N,0}$.
To simplify, this critical density is defined in terms of the present
value of Newton's gravitational constant (\ref{G_N}), $G_{N,0} =
G_*/F_0$, instead of the effective gravitational constant (\ref{Geff})
actually measured in Cavendish-type experiments. Indeed, solar-system
experiments tell us that their present values differ by less than
$0.02 \%$, as discussed in Sec.~III. [Note in passing that by
changing the value of $G_*$, one can always set $F_0 = 1$ without
loss of generality.]

In conclusion, we are left with a non-homogeneous second
order differential equation for the function $F(z)$, a situation
very different from that prevailing in general relativity. However,
the right-hand side also depends on the unknown potential $U(z)$, so
that this equation does not suffice to fully reconstruct the
microscopic Lagrangian of the theory. As we will show in Sec.~VI
below, it can nevertheless be used for a systematic study of several
scalar-tensor models, provided one of the two unknown functions is
given (or a functional dependence between them is assumed). This can
be useful as we do not expect a simultaneous release of data yielding
$H(z)$ and $\delta_m(z)$. We will see that such a study already
yields powerful constraints on the family of theories which are
viable.

On the other hand, if $\delta_m(z)$ is also experimentally
determined, and if we assume a spatially flat FRW universe
($\Omega_\kappa = 0$), we will see in the next subsection (V.B) that
the value of $\Omega_{m,0}$ as well as the function $F(z)$ can
be obtained independently of $U(z)$. Equation (\ref{F}) then
gives $U(z)$ in an {\it algebraic\/} way from our knowledge of $H(z)$,
$F(z)$ and $\Omega_{m,0}$.

Let us now assume that both $F(z)$ and $U(z)$ are known, either
because one of them was given from theoretical naturalness
assumptions, or because $\delta_m(z)$ has been experimentally
determined with sufficient accuracy. We will also assume that both
$\Omega_{m,0}$ and $\Omega_{\kappa,0}$ are known. It is then
straightforward to reconstruct the various functions of $\Phi$
entering the microscopic Lagrangian (\ref{S_JF}). In the Brans-Dicke
representation, one has $F = \Phi$, therefore the knowledge of $F(z)$
and $U(z)$ suffices to reconstruct the potential $U(\Phi)$ in a
parametric way. However, to fully determine the theory, one also
needs to know $\omega(\Phi) = \Phi Z(\Phi)$, or equivalently an
equation giving the $z$-dependence of $Z$. On the other hand, in the
simpler representation $Z = 1$ and $F(\Phi)$ unknown, we need an
equation giving the $z$-dependence of $\Phi$ to reconstruct $F(\Phi)$
and $U(\Phi)$ parametrically. These two cases, as well as any other
possible parametrization of the theory, are solved thanks to
Eq.~(\ref{dotH}) above, which reads in function of the redshift
\begin{eqnarray}
Z~\Phi'^2 &=& - F'' - \left[(\ln H)' + \frac{2}{1+z}\right]~F' +
2\left[\frac{(\ln H)'}{1+z} - \left({H_0\over
H}\right)^2\Omega_{\kappa,0}\right]~F
\nonumber\\
&& - 3(1+z)\left({H_0\over H}\right)^2 F_0 \Omega_{m,0}\ ,
\label{Phi}
\end{eqnarray}
\noindent
or equivalently
\begin{equation}
{1\over 2}Z~\Phi'^2 = -{3 F'\over 1+z} + {3 F\over (1+z)^2}
- 3 F \left({H_0\over H}\right)^2\Omega_{\kappa,0}
-{U\over (1+z)^2 H^2}
-3 (1+z) \left({H_0\over H}\right)^2 F_0 \Omega_{m,0}\ .
\label{PhiBis}
\end{equation}
In the $Z = 1$ representation, $\Phi(z)-\Phi_0$ is thus obtained by
a simple integration. In the Brans-Dicke representation, on the other
hand, $\omega(z)$ is given by an {\it algebraic\/} equation in
terms of $H(z)$, $F(z) = \Phi(z)$, and their derivatives.

It is rather obvious but anyway important to note that if the
microscopic Lagrangian (\ref{S_JF}) can be reconstructed in the
JF, it can also be directly obtained in the EF, Eq.~(\ref{S_EF}).
This allows us to check the mathematical consistency of the theory,
and notably if the helicity-0 degree of freedom $\varphi$ always
carries positive energy. One can also verify that the function
$A(\varphi)$ defining the coupling of matter to the scalar field is
well defined, and notably single valued. Finally, the second
derivative of the potential $V(\varphi)$ also gives us the sign of
the square of the scalar mass, and negative values would strongly
indicate an instability of the model. These important features cannot
easily be checked in the JF, because the sign of $Z(\Phi)$ in
Eq.~(\ref{S_JF}) is not directly related to the positivity of the
scalar-field energy (see below), and also because the second
derivative of $U(\Phi)$ does not give the precise value of its
squared mass. [As shown by Eq.~(\ref{V}), the helicity-0 degree of
freedom $\varphi$ may have a mass, $d^2V(\varphi)/d\varphi^2 \neq 0$,
even if $U(\Phi)$ is strictly constant, provided $F(\Phi)$ varies.]

Let us thus assume that $H(z)$, $\Omega_{m,0}$ and
$\Omega_{\kappa,0}$ are known, and that $F(z)$ and $U(z)$ were
reconstructed as above. Equation (\ref{A}) then gives $A(z) =
F^{-1/2}(z)$, {\it i.e.}, the Einstein-frame coupling factor
$A$ as a function of the {\it Jordan-frame\/} redshift $z$
(which is the redshift we observe).
Combining now Eq.~(\ref{varphi}) with (\ref{Phi}), we get
\begin{mathletters}
\begin{eqnarray}
\left({d\varphi\over d z}\right)^2
&=& {3\over 4}\left({F'\over
F}\right)^2 + {Z \Phi'^2\over 2F}
\label{positiveEnergy}\\
&=&
{3\over 4}\left({F'\over
F}\right)^2 - {F''\over 2 F}
-\left[{1\over 2}(\ln H)' + {1\over 1+z}\right]{F'\over F}
+{(\ln H)' \over 1+z}\nonumber\\
&&
-\left({H_0\over H}\right)^2\Omega_{\kappa,0}
-{3\over 2}(1+z)\left({H_0\over H}\right)^2 {F_0\over F}\,
\Omega_{m,0}\ ,
\label{varphiprime}
\end{eqnarray}
\label{5.5}
\end{mathletters}
or also, not eliminating the potential $U$
\begin{eqnarray}
\left({d\varphi\over d z}\right)^2
&=&{3\over 4}\left({F'\over
F}\right)^2 -{3 F'\over (1+z) F}
+{3\over (1+z)^2}
-3 \left({H_0\over H}\right)^2\Omega_{\kappa,0}
\nonumber\\
&&- {U\over (1+z)^2 F H^2}
- 3 (1+z)\left({H_0\over H}\right)^2 {F_0\over F}\,
\Omega_{m,0}\ .
\label{varphiprimeBis}
\end{eqnarray}
\noindent
The EF scalar $\varphi$ is thus also known as a
function of the {\it Jordan-frame\/} redshift $z$ (up to an
additive constant $\varphi_0$ which can be chosen to vanish without
loss of generality), and one can reconstruct $A(\varphi)$ in a
parametric way. Similarly, the EF potential $V(\varphi)$,
Eq.~(\ref{V}), can be reconstructed from our knowledge of $F(z)$,
$U(z)$ and $\varphi(z)$.

Since $\varphi$ describes the actual helicity-0 degree of freedom of
the theory, this field must carry only positive energy excitations,
and $(d\varphi/dz)^2$ must be positive. On the other
hand, the tensor and scalar degrees of freedom are mixed in the JF,
and the positivity of energy does not imply that $Z \Phi'^2$
should always be positive. Actually, Eq.~(\ref{positiveEnergy}) shows
that it can become negative when ${3\over 4}(\ln F)'^2$ happens to be
larger than $\varphi'^2$, which can occur in perfectly regular
situations. [We will see an explicit example in Sec.~VI.A below.]
This underlines that the parametrization $Z = 1$ can sometimes be
singular: The derivatives of $\Phi$ may become purely imaginary
although the scalar degree of freedom $\varphi$ is well defined. On
the other hand, the Brans-Dicke representation is well behaved
($\Phi'^2$ remains always positive), and the positivity of energy
simply implies the well-known inequality $\omega(\Phi) \geq -{3\over
2}$. Actually, the particular value $\omega = -{3\over 2}$ is also
singular, as it corresponds to an infinite coupling strength $\alpha =
(2\omega+3)^{-1/2}$ between matter and the helicity-0 degree of
freedom $\varphi$. The domain for which the $Z=1$ parametrization is
pathological although the theory remains consistent simply
corresponds to $-{3\over2} < \omega(\Phi) < 0$, or $|\alpha| >
1/\sqrt{3}$.

\subsection{Perturbations}
Although the perturbations will not be used in Sec.~VI below, we
emphasize that the phenomenological reconstruction of the full
microscopic Lagrangian can be implemented without any ambiguity if
fluctuations are taken into account. For completeness, we review now
this part of our program. We assume that both $H(z)$ and the matter
density perturbation $\delta_m(z)$ are
experimentally determined with enough accuracy, and as in Sec.~IV.B
above, we focus our discussion on the case of a spatially flat FRW
universe ($\Omega_\kappa = 0$). We also assume that matter is
dustlike ($p=0$), and the perturbation equations of Sec.~IV.B are
thus simplified. In particular, Eq.~(\ref{conserv_z}) reduces to the
mere identity $\phi = \dot v$.

We consider comoving wavelengths $\lambda \equiv a/k$ much shorter
(for recent times) than the Hubble radius $H^{-1}$, and also shorter
than the inverse mass of the scalar field:
\begin{equation}
k^2/ a^2 \gg \max\left(H^2, A^{-2} |d^2 V/ d\varphi^2|\right)\ .
\label{largek}
\end{equation}
Two different reasonings can now be used to reach the same
conclusions. The first one, explained in Ref.~\cite{beps00},
consists in taking the formal limit $k\rightarrow\infty$ in the
various perturbation equations. Then, the leading terms are either
those containing $\delta_m$ or those multiplied by the large factor
$k^2 / a^2$. One also needs to consider only the growing adiabatic
mode of Eq.~(\ref{fluctPhi}), for which $|\ddot{\delta \Phi}|\ll
k^2a^{-2}|\delta\Phi|$.

The other reasoning needs a simpler (but {\it a priori\/} stronger)
hypothesis. One assumes that the logarithmic time derivative of
any quantity, say $f$, is at most of order $H$~: $|\dot f| \lesssim
|H f|$. Physically, this means that the expansion of the universe is
driving the time evolution of every physical quantity. Then the
hypothesis $k^2/ a^2 \gg H^2$ suffices to derive straightforwardly
all the following approximations.

Note that both reasonings correspond in fact to the same physical
situation of a weakly-coupled light scalar field. In the case of a
strongly-coupled but very massive scalar (see the second paragraph
of Sec.~III), the equations cannot be approximated as shown below,
and the time evolution of density fluctuations does not follow the
same law. For instance, in the particular model considered in
Ref.~\cite{fiz99}, one always finds a strong clustering of the
scalar field at small scales. Indeed, this model corresponds to the
choice $F = \Phi$ and $Z = 0$ in action (\ref{S_JF}), and
Eq.~(\ref{fluctPhi}) can then be rewritten as
$(d^2U/d\Phi^2)\delta\Phi = (k^2/a^2)(\phi-2\psi) -
3(\ddot\psi+4H\dot\psi+H\dot\phi) - 2\phi(dU/d\Phi)$. Therefore,
even if the scalar field is very massive ($d^2U/d\Phi^2$ large), one
finds that it is anyway strongly clustered for comoving wavelengths
$a/k$ shorter than the inverse mass, {\it i.e.}, in the formal limit
$k\rightarrow\infty$. Although this is {\it a priori\/} not
forbidden by observations of gravitational clustering, since the
inverse mass must be much smaller than the astronomical unit in this
model, this is anyway an indication of its probable instability.
We will not consider such heavy scalar fields any longer in this
paper, and we now come back to the class of weakly-coupled
light-scalar models, which are the most natural alternatives to
general relativity.

Setting $B \equiv \psi + H v$ and making use of (\ref{conserv_z}),
one can write (\ref{conserv_0}) as
\begin{equation}
\ddot \delta_m + 2H\dot \delta_m + {k^2\over a^2}\,\phi
= 3\ddot B+6H\dot B \approx 0~,
\label{ddotdeltam}
\end{equation}
where the right-hand side is negligible with respect to each
separate term of the left-hand side because of the above
hypotheses. Note that (\ref{ddotdeltam}) just reproduces the
standard evolution equation for matter perturbations. Using
(\ref{fluctPhiZ1}), we also arrive at
\begin{equation}
\delta \Phi \approx (\phi - 2\psi)~{dF/d\Phi\over Z} \approx
- \phi~{F~dF/d\Phi\over ZF+2(dF/d\Phi)^2}~,
\label{dPhi}
\end{equation}
where the second equality is a consequence of Eq.~(\ref{psi}).
In the case of GR plus a minimally coupled scalar field, one finds
that $\delta\Phi\propto k^{-2}\phi$ in the limit
$k\rightarrow\infty$, so that the scalar field is not
gravitationally clustered at small scales \cite{St98}. This is in
agreement with the observational fact that the dark matter described
by the $\Lambda$-term should remain unclustered up to comoving scales
$R\sim 10\, h^{-1}(1+z)^{-1}$ Mpc (where we recall that $h^{-1} \equiv
100\, H_0^{-1}\ {\rm km}\ {\rm s}^{-1}\ {\rm Mpc}^{-1}$). On the
other hand, in our scalar-tensor framework, Eq.~(\ref{dPhi}) shows
that the scalar field {\it is\/} clustered at arbitrarily small
scales, but only weakly because the derivative $|dF/d\Phi|$ is
experimentally known to be small [see the solar-system constraint
(\ref{dF0}), and the limit
$\alpha^2 \lesssim 0.1$ justified in \cite{beps00} for redshifts $z
\lesssim 1$]. The class of models we are considering, involving a
light scalar field weakly coupled to matter, is thus also in agreement
with observations of gravitational clustering.

Finally, still under the above hypotheses, Eq.~(\ref{fluctmetr})
implies
\begin{equation}
- 2 {k^2\over a^2}F\phi \approx 8\pi G_* \rho~\delta_m
+ {k^2\over a^2}\frac{dF}{d\Phi} \delta \Phi~.
\label{Poi}
\end{equation}
Remembering the definition (\ref{Geff}) for $G_{\rm eff}$, and
using (\ref{dPhi}) above, Eq.~(\ref{Poi}) can be recast in a form
which exhibits its physical content:
\begin{equation}
{k^2\over a^2}\phi \approx - 4 \pi G_{\rm eff} \rho\delta_m~.
\label{Poi2}
\end{equation}
Poisson's equation is thus simply modified by the substitution of
Newton's constant $G$ by $G_{\rm eff}$, the effective gravitational
constant between two close test masses! This conclusion was also
reached in \cite{GaLo00}, but only for Brans-Dicke theory with a
constant parameter $\omega$, while we have derived it for an
arbitrary (light) scalar-tensor theory. As discussed in Sec.~III
above, expression (\ref{Geff}) is valid only if the distance between
the test masses is negligible with respect to the inverse scalar
mass. The physical reason why this expression appears in Poisson's
equation (\ref{Poi2}) is just that we are working in the short
wavelength limit (\ref{largek}): The frequency of the waves we are
considering is so large that the scalar field behaves as if it were
massless.

Combining (\ref{ddotdeltam}) with (\ref{Poi2}), we now arrive at
our final evolution equation for $\delta_m$~:
\begin{equation}
{\ddot \delta_m} + 2H {\dot \delta_m} - 4\pi G_{\rm eff}\,
\rho~\delta_m\approx 0~.
\label{deltam}
\end{equation}
In terms of the redshift $z$, this reads
\begin{equation}
H^2~\delta_m'' + \left(\frac{(H^2)'}{2} -
{H^2\over 1+z}\right)\delta_m'
\approx {3\over 2} (1+z) H_0^2 {G_{\rm eff}(z)\over
G_{N,0}}~\Omega_{m,0}~\delta_m~.
\label{delz}
\end{equation}
\noindent
Provided we can extract from observation both physical
quantities $H(z)$ and $\delta_m(z)$ with sufficient accuracy, the
explicit reconstruction of the microscopic Lagrangian is obtained in
the following way. Starting from (\ref{delz}) and using the fact
that {\it today} $G_{\rm eff,0}$ and $G_{N,0}$ differ by less than
$0.02 \%$, Eq.~(\ref{delz}) evaluated at present gives us the
cosmological parameter $\Omega_{m,0}$ with the same accuracy. Then,
returning to Eq.~(\ref{delz}) for arbitrary $z$, we get
$G_{\rm eff}(z)=p(z)$, where $p(z)$ is a known function of the
observables $H(z)$, $\delta_m(z)$, and their derivatives.
Using now Eq.~(\ref{Phi}) and expression (\ref{Geff}) for $G_{\rm
eff}$, we get a nonlinear second order differential equation
for $F(z)$, which can be solved for given $F_0$ and $F'_0$ [one can
always set $F_0 = 1$ without loss of generality, while
$F'_0$ is constrained by Eq.~(\ref{dF0})]. After we have found
$F(z)$, we can plug it into (\ref{F}) to determine $U(z)$ in an
algebraic way. The final step is explained in the previous
subsection, above Eq.~(\ref{Phi}), for the various possible
parametrizations of action (\ref{S_JF}): In the $Z=1$
parametrization, $\Phi(z)-\Phi_0$ is obtained by a simple
integration of Eq.~(\ref{Phi}), while in the Brans-Dicke
parametrization ($F(\Phi) = \Phi$), $\omega(z)$ is given
algebraically by the same Eq.~(\ref{Phi}). This enables us to
reconstruct $F(\Phi)$ (or $\omega(\Phi)$) and $U(\Phi)$ as functions
of $\Phi-\Phi_0$ for that range corresponding to the data.

Actually, for sufficiently low redshifts $z \lesssim 1$,
Eq.~(\ref{delz}) can be simplified without losing too much
accuracy. Indeed, as shown in Ref.~\cite{beps00}, the square of the
matter-scalar coupling strength $\alpha$, Eq.~(\ref{alpha}), is at
most of order $10 \%$ for such redshifts. Moreover, under natural
assumptions, much smaller values of $\alpha^2$ are generically
predicted in scalar-tensor theories \cite{dn93,w98}. Therefore,
$G_{\rm eff}$ and $G_N$ differ by less than $\sim 10 \%$ for
redshifts $z \lesssim 1$, and Eq.~(\ref{delz}) can be used to obtain
$G_{\rm eff} / G_{N,0} \approx G_N / G_{N,0} = F_0/F$ with the same
accuracy. The interest of this simplification is that $F(z)$ is now
given by an {\it algebraic\/} equation. In the Brans-Dicke
representation, {\it all\/} the steps of the reconstruction program
are thus algebraic, Eq.~(\ref{F}) giving $U(z)$, and Eq.~(\ref{Phi})
giving $\omega(z)$. The only non-algebraic step is the final
parametric reconstruction of $U(\Phi)$ and $\omega(\Phi)$.

Let us end this section by a few comments on the observational
accuracy which will be needed for this reconstruction program to
be implemented. First, Eq.~(\ref{delz}) allows to reconstruct
$F(z)$ only if $\delta_m'$ and $\delta_m''$ are both determined
with enough accuracy. Moreover, the second derivative of this
reconstructed $F(z)$ is needed in Eq.~(\ref{F}) to obtain
$U(z)$. Therefore, the actual reconstruction of the potential
depends {\it a priori\/} on the fourth derivative of $\delta_m(z)$,
so that extremely clean data seem to be necessary. However, the
situation is better than this naive derivative counting suggests.
Indeed, the above estimates for $\alpha^2$ show that $F(z)$ does
not vary much on the redshift interval $0\leq z \lesssim 1$.
Therefore, the first two terms of Eq.~(\ref{F}), involving $F'$ and
$F''$, are expected to be negligible with respect to the third one
involving $F$. A noisy experimental determination of
$\delta_m'''(z)$ and $\delta_m''''(z)$ is thus not a serious
difficulty for our reconstruction program. On the other hand,
clean enough data are still needed to determine $F(z)$ from
Eq.~(\ref{delz}), using $\delta_m(z)$ and its first two
derivatives. Before such clean data are available, it will be
sufficient to verify that Eq.~(\ref{delz}) is consistent with a
slowly varying $F(z)$. In the next section, we will show that
interesting theoretical constraints can anyway be obtained without
knowing at all the density fluctuation $\delta_m(z)$, but using
only the luminosity distance $D_L(z)$ and consistency arguments
within particular subclasses of scalar-tensor models.

\section{Constraints from an accelerating universe}
In Ref.~\cite{SRSS}, a fit of presently known supernovae events
has been performed to obtain the luminosity distance $D_L(z)$ up to
redshifts $z\sim 1$, of course still with large uncertainties.
Although this is not yet sufficient to constrain seriously
scalar-tensor models, we can expect clean data on $D_L(z)$ in the
near future from additional supernovae events, and anyway earlier
than for the density perturbations $\delta_m(z)$. The SNAP satellite
will in particular observe thousands supernovae events up to
$z\approx 1.7$. In this section, we will concentrate on the
theoretical constraints that can be extracted from the knowledge of
$D_L(z)$ alone, and therefore of $H(z)$ using Eq.~(\ref{HvsDL}).
We will thus only use the results of subsection V.A above. Since the
knowledge of this function does not suffice to fully reconstruct the
microscopic Lagrangian (\ref{S_JF}), we will need additional
assumptions on one of the functions it involves, either $F$ (or $Z$,
depending on the parametrization) or the potential $U$. One may also
assume a functional relation between $F$ and $U$ (for instance
$U\propto F^M$ as in Ref.~\cite{Am99}).

To emphasize as clearly as possible what kind of constraints can be
imposed on scalar-tensor theories, we shall consider
the worst situation for them. Let us assume that the observed
function $H(z)$ will be exactly given by Eq.~(\ref{H2}) for
$\kappa = 0$, $F=\Phi=1$, and $U=\Lambda \equiv
3H_0^2\Omega_{\Lambda,0}$~:
\begin{equation}
(H/H_0)^2 = \Omega_{\Lambda,0} + \Omega_{m,0} (1+z)^3~.
\label{H2GR}
\end{equation}
Of course, such an observation would {\it a priori\/} call for the
following standard interpretation: Gravity is correctly described by
general relativity, and we live in a flat universe filled with
dustlike matter and a cosmological constant, with corresponding
present energy densities (relative to the critical density)
$\Omega_{m,0}$ and $\Omega_{\Lambda,0}$. However, for our purpose,
Eq.~(\ref{H2GR}) should just be considered as kinematical. It tells
us how the universe expands with redshift $z$, but we are free to
assume that the dynamics of the expansion is governed by a
scalar-tensor theory. Therefore, $\Omega_{m,0}$ and
$\Omega_{\Lambda,0}$ are here mere parameters, whose names refer
to their physical significance in the framework of GR. Of course,
one should not forget that they do not have the same interpretation
within scalar-tensor theories.

For our numerical applications, we will further take the present
estimates based on combined CMB fluctuations and supernovae
observations (they will be determined more accurately
by future experiments):
\begin{equation}
\Omega_{\Lambda,0} \approx 0.7~,\quad \Omega_{m,0} \approx 0.3~.
\label{OmegaValues}
\end{equation}
For these numerical values, (\ref{H2GR}) is consistent with the
presently available luminosity distance $D_L(z)$ up to $z\sim 1$.
Actually the best-fit universe, if we assume flatness, gives
$\Omega_{\Lambda,0}=0.72$ and $\Omega_{m,0}=0.28$. We have chosen to
work directly with the exact form (\ref{H2GR}), instead of
the $D_L(z)$ extracted from observation, in order to clarify the
physical content of our results. Indeed, the present observational
estimates for $D_L(z)$ are still too imprecise to constrain strongly
the class of scalar-tensor theories we are considering. Moreover,
some of our results below depend crucially on the fact that $H(z)$
keeps the form (\ref{H2GR}) up to redshifts $z\sim 2$, which have not
yet been reached experimentally. To relate our results to those
obtained in \cite{SRSS,WA} using fitting functions or an expansion in
powers of $z$, one just needs to use Eq.~(\ref{HvsDL}): Our exact
expression (\ref{H2GR}) for $H(z)$ corresponds to some exact
expression for
$D_L(z)$.

To summarize, we are assuming in this section that future observations
of the luminosity distance $D_L(z)$ will provide a $H(z)$ of the form
(\ref{H2GR}) with the numerical values (\ref{OmegaValues}). This
implies notably that our Universe is presently accelerating. On the
other hand, we are {\it not\/} assuming that the correct theory of
gravity is necessarily GR plus a cosmological constant. The main
question that we will address is therefore the following: Would such an
``observed'' $H(z)$ necessarily rule out the existence of a scalar
partner to the graviton? If not, would it be possible to reproduce
(\ref{H2GR}) within a more natural scalar-tensor theory, in which
$\Omega_{\Lambda,0}$ could be explained by a ``generalized
quintessence'' mechanism?

We will first analyze in subsection A the simplest subclass of
scalar-tensor theories that we can consider, namely when $U=0$ in
action (\ref{S_JF}). Since this is {\it a priori\/} the subclass
which differs the most from GR plus a cosmological constant, this
study will be rather detailed, and it will allow us to underline the
mathematical and physical meaning of the constraints that are
obtained. Subsection B will be again devoted to the case of a
massless scalar field, but combined with a cosmological constant.
As its conclusions basically confirm those of subsection A, we
will present them more concisely. Finally, we will briefly discuss
in subsection C the cases where one imposes particular forms
for the coupling function $F$ in action (\ref{S_JF}), and one
reconstructs the potential $U$ from the background equations
(\ref{F})--(\ref{Phi}). The case of a given functional dependence
between $F$ and $U$ will also be addressed.

\subsection{Case of a vanishing scalar-field potential}
Since a cosmological constant can be interpreted as a particular
case of scalar-field potential, it is instructive to analyze
whether an observed expansion like (\ref{H2GR}) could be reproduced
in a theory {\it without\/} any potential, and we now study
Eqs.~(\ref{F})--(\ref{5.5}) for $U(\Phi)=0=V(\varphi)$.
This case can be analyzed using the second order differential
equation (\ref{F}) for $F$, which simplifies significantly if one
introduces a function $f$ such that
\begin{equation}
F(z)/F_0 \equiv (1+z)^2 f(1+z)~.
\label{smallf}
\end{equation}
[As mentioned in Sec.~V.A above, one can also set $F_0 = 1$ without
loss of generality.] Then, using the assumed ``experimental''
expression (\ref{H2GR}) for $H(z)$, and writing (\ref{F}) in terms
of $x\equiv 1+z$, we get
\begin{equation}
(\Omega_{\Lambda,0}+\Omega_{m,0}\, x^3) x f''(x)
+ {3\over 2}\Omega_{m,0}\, x^3 f'(x)
-4 \Omega_{\kappa,0}\, x f(x)
= 3 \Omega_{m,0}~.
\label{Eqf}
\end{equation}
To avoid any confusion, let us recall that $\Omega_{\Lambda,0}$ (and
the two occurrences of $\Omega_{m,0}$ in the left-hand side) comes from
the ``observed'' cosmological function (\ref{H2GR}), notwithstanding
the fact that there is {\it no\/} cosmological constant in the model we
are considering. The value $\Omega_{m,0}$ appearing in the right-hand
side stands for the present relative energy density of dustlike matter.
We assume that it takes the same numerical value (\ref{OmegaValues}) as
in the ``observed'' $H(z)$ (\ref{H2GR}). Equation (\ref{Eqf}) tells us
how we should choose $f(x)$ to mimic exactly this $H(z)$ in the present
potential-free theory. In other words, $\Omega_{\Lambda,0}$ and
$\Omega_{m,0}$ are two numbers assumed to be given by experiment, and
we wish to fit $f(x)$ and $\Omega_{\kappa,0}$ to satisfy
Eq.~(\ref{Eqf}).

To integrate this second-order differential equation, we need two
initial conditions for $f$ and its derivative. The first one is
an obvious consequence of Eq.~(\ref{smallf}) taken at $z = 0$, and
we simply get $f(1) = 1$. The second one should be such that the
solar-system bound (\ref{dF0}) is satisfied. For instance, if
$\Phi'_0$ does not vanish, it is {\it sufficient\/} to impose
$F'_0=0$, {\it i.e.}, $f'(1) = -2$ using Eq.~(\ref{smallf}).
This corresponds to a scalar-tensor theory which has been attracted
towards an extremum of $F$ during the cosmological expansion of the
universe ({\it cf.} \cite{dn93,w98}), so that it is presently
strictly indistinguishable from general relativity in solar-system
experiments. [The full allowed domain for $f'(1)$ will be explored
below in a numerical way.]

\subsubsection{Spatially flat universe}
We consider first our potential-free model in a spatially flat FRW
universe ($\Omega_{\kappa,0} = 0$). Then Eq.~(\ref{Eqf}) becomes a
first-order differential equation for $f'$, and its integration
yields
\begin{equation}
f'(x) = {1\over \sqrt{1+\zeta
x^3}}\left[\zeta\ln\left(
{\sqrt{1+\zeta}+1\over\sqrt{1+\zeta x^3}+1}
\cdot
{\sqrt{1+\zeta x^3}-1\over\sqrt{1+\zeta}-1}\right)
-2\sqrt{1+\zeta}\right]\,,
\label{fprime}
\end{equation}
where we have set $\zeta\equiv \Omega_{m,0}/\Omega_{\Lambda,0}$, and
where the final constant inside the square brackets has been chosen
to impose $f'(1) = -2$ ({\it i.e.}, $F'_0 = 0$). The function $f(x) =
1+ \int_1^x{f'(y)dy}$ can be explicitly written in terms of
generalized hypergeometric functions, but its complicated expression
will not be useful for our purpose. Let us just quote the first
order of its expansion in powers of
$\Omega_{m,0}/\Omega_{\Lambda,0}$~:
\begin{equation}
f(x) = 3 - 2 x +
{1\over 4}(15 - 16 x + x^4 + 12 x \ln x){\Omega_{m,0}\over
\Omega_{\Lambda,0}}
+O\left({\Omega_{m,0}^2\over\Omega_{\Lambda,0}^2}\right)\,.
\label{fexpanded}
\end{equation}
In conclusion, Eq.~(\ref{Eqf}) could be integrated analytically,
in the particular case of a spatially flat universe. This means that
at least in the vicinity of $z = 0$, there {\it a priori\/} exists a
potential-free scalar-tensor theory which exactly mimics general
relativity plus a cosmological constant.

However, the theory is mathematically consistent only if $F(z)$
remains strictly positive. [If $F$ vanishes, then the coupling
function $A(\varphi)$, Eq.~(\ref{A}), between matter and the
helicity-0 degree of freedom $\varphi$ diverges, and if $F$ becomes
negative, the graviton carries negative energy.] Let us thus compute
the value $z_{\rm max}$ for which $F(z_{\rm max})$, or $f(1+z_{\rm
max})$, vanishes for the first time. Because of the complexity of the
solution $f$, we did not find a close analytical expression for
$z_{\rm max}$, but its expansion in powers of
$\Omega_{m,0}/\Omega_{\Lambda,0}$ can be obtained straightforwardly:
\begin{equation}
z_{\rm max} = {1\over 2}
+ {9\over 4}\left(\ln{3\over 2} - {7\over 32}\right)
{\Omega_{m,0}\over\Omega_{\Lambda,0}}
+ {3\over 8}\left[-{5105\over 1792}
+ {21\over 8}\,\ln{3\over 2}
+ 9\left(\ln{3\over 2}\right)^2
\right]
{\Omega_{m,0}^2\over\Omega_{\Lambda,0}^2}
+O\left({\Omega_{m,0}^3\over\Omega_{\Lambda,0}^3}\right).
\label{zmax}
\end{equation}
Numerically, for the values (\ref{OmegaValues}) of $\Omega_{\Lambda,0}$
and $\Omega_{m,0}$, we find $z_{\rm max} \approx 0.66$. In conclusion,
this scalar-tensor model is able to mimic general relativity plus a
cosmological constant, but only on the small interval $z \leq 0.66$. If
future observations of type Ia supernovae give a behavior of $H(z)$ of
the form (\ref{H2GR}) on a larger interval, say up to $z\sim 1$, then
the present scalar-tensor theory will be ruled out. This example of a
vanishing potential illustrates a conclusion that we will reobtain
below for more general theories: The determination of the form of
$H(z)$ over some (even rather small) redshift interval is in fact more
constraining than the precise value of the parameters
$\Omega_{m,0},~\Omega_{\Lambda,0}$ themselves. Indeed, Eq.~(\ref{zmax})
clearly shows that $z_{\rm max}$ cannot exceed 1 even in the presumably
unrealistic case of $\Omega_{m,0}\approx \Omega_{\Lambda,0}$. [A
calculation using the exact expression for $f(x)$ shows that $z_{\rm
max}$ would exceed 1 only for $\Omega_{m,0}/\Omega_{\Lambda,0} \geq
1.59$.] Note that all the results obtained are independent of the
parameter $H_0$.

\subsubsection{Spatially curved universe}
One could try to increase $z_{\rm max}$ by considering a spatially
curved FRW universe. We did not solve Eq.~(\ref{Eqf}) in the most
general case, but since we wish to compute the corrections to
Eq.~(\ref{zmax}) due to a small value of
$|\Omega_{\kappa,0}/\Omega_{\Lambda,0}|$, it is sufficient to
work at zeroth order in $\Omega_{m,0}/\Omega_{\Lambda,0}$.
Let us thus set $\Omega_{m,0} = 0$ in Eq.~(\ref{Eqf}), which
reduces to
\begin{equation}
\Omega_{\Lambda,0} f''(x) -4 \Omega_{\kappa,0} f(x) = 0~.
\label{EqfOm0}
\end{equation}
Its solution is obviously a sine if $\Omega_{\kappa,0} < 0$ ({\it
i.e.}, $\kappa = +1$, closed universe), or a hyperbolic sine for
$\Omega_{\kappa,0} > 0$ ({\it i.e.}, $\kappa = -1$, open universe).
Taking into account the initial conditions $f(1) = 1$ and $f'(1) =
-2$, we thus get
\begin{mathletters}
\begin{eqnarray}
f(1+z) = \cos(2\xi z) - {1\over \xi}\sin(2\xi z)\qquad &{\rm for}&
\quad \xi^2\equiv -{\Omega_{\kappa,0}\over\Omega_{\Lambda,0}} > 0~,
\label{fcos}\\
f(1+z) = \cosh(2\xi z) - {1\over \xi}\sinh(2\xi z)\qquad &{\rm
for}&
\quad \xi^2\equiv +{\Omega_{\kappa,0}\over\Omega_{\Lambda,0}} > 0~.
\label{fcosh}
\end{eqnarray}
\label{fOk0}
\end{mathletters}
The first zero of $f(1+z)$ is then reached either at $z_{\rm max} =
{1\over 2\xi}\arctan\xi$ or at ${1\over 2\xi}\,{\rm arctanh}\,\xi$.
In both cases, the expansion in powers of $\xi$ gives $z_{\rm max}
\approx {1\over 2} + {1\over
6}\,\Omega_{\kappa,0}/\Omega_{\Lambda,0}$. Working perturbatively,
one can also compute the correction to this expression due to the
nonzero value of $\Omega_{m,0}$, and one finds that $z_{\rm max}$ is
given by Eq.~(\ref{zmax}) above plus the following correction:
\begin{equation}
\delta z_{\rm max} =
{1\over 6}\left[
1 + { {456\ln(3/2) - 163} \over 16} \,
{\Omega_{m,0}\over\Omega_{\Lambda,0}}
+O\left({\Omega_{m,0}^2\over
\Omega_{\Lambda,0}^2}\right)
\right]
{\Omega_{\kappa,0}\over\Omega_{\Lambda,0}}
+O\left({\Omega_{\kappa,0}^2\over\Omega_{\Lambda,0}^2}\right)\,.
\label{deltazmax}
\end{equation}
In conclusion, $z_{\rm max}$ can be slightly enlarged if we consider
our potential-free scalar-tensor theory in an open FRW universe
($\Omega_{\kappa,0} > 0$). Numerically, for the values
(\ref{OmegaValues}) of $\Omega_{\Lambda,0}$ and $\Omega_{m,0}$,
we find $\delta z_{\rm max} \approx 0.26\,
\Omega_{\kappa,0}/\Omega_{\Lambda,0}$. Since the latest
experimental data on CMB temperature fluctuations already constrain
$|\Omega_{\kappa,0}|$ to be small (see the latest Boomerang and
Maxima data), and actually an open universe is unlikely while a
marginally closed universe is still acceptable, we thus recover the
same qualitative conclusion as in the spatially flat case: It is
possible to mimic general relativity plus a cosmological constant
within a potential-free scalar-tensor theory only on a small redshift
interval $z \lesssim 0.8$.

\subsubsection{Numerical integrations}
The above conclusions have been confirmed by numerical integrations of
Eqs.~(\ref{F})--(\ref{5.5}), still assuming a Hubble diagram consistent
with (\ref{H2GR}). Instead of considering only theories which are
presently indistinguishable from general relativity ($F'_0 = 0$), we
imposed arbitrary initial conditions for $F'$, and computed the
corresponding value of the present scalar-matter coupling strength
$\alpha_0$, Eq.~(\ref{alpha}). In the case of a spatially flat FRW
universe, we recovered that the solar-system bound (\ref{dF0}) imposes
the limit $z_{\rm max} \approx 0.68$, consistently with the above
analytical estimate (\ref{zmax}). In other words, the constraint
(\ref{dF0}) is so tight that even taking the largest allowed value for
$|\alpha_0|$ does not change significantly $z_{\rm max}$. Figure 1
displays the reconstructed $F(z)$ for this maximal $|\alpha_0|$, and
one can note that its slope at $z=0$ is visually indistinguishable from
the horizontal. This figure also plots the Einstein-frame scalar
$\varphi$, Eq.~(\ref{varphi}), which is the actual helicity-0 degree of
freedom of the theory. Notice that it diverges at $z_{\rm max}$, so
that the theory loses its consistency beyond this value of the
redshift.

Curiously, we found that even if no experimental constraint like
(\ref{dF0}) is imposed on $|\alpha_0|$ ({\it i.e.}, even if we
forget that solar-system experiments confirm very well general
relativity), then the mathematical consistency of the theory anyway
imposes $z < 3.5$. In fact, Eq.~(\ref{F}) alone can be solved for
arbitrary large values of $z$, {\it i.e.}, there exist initial
values of $F'_0$ such that $F(z)$ remains positive for any $z$.
However, the values of $F'_0$ needed to integrate Eq.~(\ref{F})
beyond $z = 3.5$ correspond to negative values of $\alpha_0^2 =
(2\omega_0+3)^{-1}$ (where $\omega_0$ denotes the present value of
the Brans-Dicke parameter). In other words, the expression of
$(d\varphi/dz)^2$ given by Eq.~(\ref{5.5}) would become negative
around $z=0$, and the helicity-0 degree of freedom would thus need
to carry negative energy at least on a finite interval of $z$, if
one wished to integrate Eqs.~(\ref{F})--(\ref{5.5}) beyond $z =
3.5$.

Figure 2 displays the maximum redshift $z_{\rm max}$ consistent
with the positivity of energy of both the graviton and the scalar
field, but for any value of the present matter-scalar coupling
strength $|\alpha_0|$. As underlined above, one finds that $z_{\rm
max}$ can never be larger than $3.5$. This figure also indicates
the present solar system bound on $|\alpha_0|$, corresponding to
$z_{\rm max} \approx 0.68$ as in Fig.~1. The limiting case of a
vanishing $|\alpha_0|$, {\it i.e.}, of a scalar-tensor theory which
is presently {\it strictly\/} indistinguishable from GR in the solar
system, corresponds to $z_{\rm max} \approx 0.66$, as was derived
analytically in Eq.~(\ref{zmax}). Figure 2 also indicates the
range of values for $|\alpha_0|$ that are generically obtained in
Refs.~\cite{dn93} while studying the cosmological evolution of
scalar-tensor theories at earlier epochs in the matter-dominated
era: The theory is attracted towards a maximum of $F$ ({\it i.e.}, a
minimum of $\ln A(\varphi)$) so that the present value of
$|\alpha_0|$ is expected to be extremely small. Finally, this figure
also displays the maximum value of $|\alpha_0|$ for which the
parametrization $Z(\Phi)=1$ of action (\ref{S_JF}) has a meaning.
Beyond $|\alpha_0| = 1/\sqrt{3}$ ({\it i.e.}, for a Brans-Dicke
parameter $-{3\over 2} < \omega_0 < 0$), one would get $\Phi'^2_0 <
0$ in this parametrization. In other words,
Eqs.~(\ref{F})--(\ref{Phi}) cannot be integrated consistently beyond
$z \approx 1.58$ if one sets $Z(\Phi) = 1$, whereas the Brans-Dicke
or the Einstein-frame representations show that the theory can be
mathematically consistent up to $z\approx 3.5$ ($\varphi'^2$ remains
positive). This underlines that the $Z=1$ parametrization may be
sometimes pathological.

Our numerical integration of Eq.~(\ref{varphiprime}) not only allowed
us to check the positivity of the scalar field energy, but also to
reconstruct parametrically the matter-scalar coupling function
$A(\varphi)$. Since $A = F^{-1/2}$, Eq.~(\ref{A}), we know that $A(z)$
is finite and strictly positive over the interval $[0,z_{\rm max}[$,
but we also checked that it is single valued over this interval. This
means that if $\varphi(z)$ can take several times the same value for
different $z$, they must correspond also to the same value of $A(z)$.
Actually, since Eq.~(\ref{varphiprime}) does not fix the sign of
$d\varphi/dz$, one should keep in mind that $\varphi$ can oscillate
around a constant value $\varphi_{\rm min}$. If the numerical
integration confuses the two points $\varphi_{\rm min} \pm
\varepsilon$, but if $A(\varphi)$ happens not to be symmetrical
around $\varphi_{\rm min}$, it may look like a bi-valued function.
When such a situation occurred in our programs, we always verified
that a single-valued $A(\varphi)$ could be defined consistently by
unfolding it around the oscillation points of $\varphi$. Figure 3
illustrates such a situation, for an intentionally unrealistic value
of $|\alpha_0|$ in order to clarify the plots. [The value $|\alpha_0|
= 1$ is inconsistent with the solar-system bound (\ref{dF0}), but it
corresponds anyway to a mathematically consistent theory, although
the $Z=1$ parametrization cannot be used in this case.]

All the functions $\ln A(\varphi)$ that we reconstructed have
similar convex parabolic shapes. This is consistent with the results
of Refs.~\cite{dn93,w98}, showing that the scalar field is
generically attracted towards a minimum of $\ln A(\varphi)$ during
the expansion of the universe. If we had found models such that the
present epoch ($z=0$) is close to a {\it maximum\/} of $\ln A$, this
would have meant that the theory is unstable, and that we have
extremely fine tuned it to be consistent with solar-system
constraints. On the contrary, the convex functions $\ln A(\varphi)$
that we obtained show that these scalar-tensor models are
cosmologically stable, {\it i.e.}, that the tight bounds (\ref{dF0})
are in fact natural consequences of the attractor mechanism described
in \cite{dn93,w98}.

We have checked that reducing the parameter $\Omega_{\Lambda,0}$
allows us to extend the integration region in the past,
consistently with the above analytical results. For instance, when
we vary $\Omega_{\Lambda,0}$, still satisfying $\Omega_{m,0} = 1 -
\Omega_{\Lambda,0}$ and setting $\Omega_{\kappa,0} = 0$, we find
that $\Omega_{\Lambda,0} < 0.02$ is required in order to
integrate the equations up to a redshift $z=5$. This would
correspond to $\Omega_{m,0}/\Omega_{\Lambda,0} > 50$, {\it i.e.},
100 times larger than present estimates.

We also added random noise to our assumed $H(z)$,
Eq.~(\ref{H2GR}), and verified that the conclusions are not changed
qualitatively provided $H(z)$ is known over a wide enough redshift
interval. This means that the experimental determination of the
luminosity distance $D_L(z)$ needs not be very precise to be quite
constraining, provided redshifts of order $z\sim 2$ are probed.
As an illustration, let us take the exact expression $H(z)$ of
Eqs.~(\ref{H2GR})--(\ref{OmegaValues}) for discrete values of the
redshift, say $z = 0, 0.1, 0.2, 0.3, \dots$, and let us add or
subtract randomly between $0$ and $30 \%$ to the corresponding
$H(z)$. Then, we may fit a polynomial through these ``noisy''
values of $H(z)$, and use our numerical programs to integrate the
background equations (\ref{F})--(\ref{5.5}) and reconstruct $F$.
We found that there always exists a maximum redshift beyond which
$F$ is negative (and the theory thus inconsistent). Figure 4
displays the two extreme values of $z_{\rm max}$ that we obtained
with hundreds of such ``deformed'' $H(z)$: It is sometimes
even smaller than for the ``exact'' $H(z)$ of Eq.~(\ref{H2GR}), and
sometimes larger but never greater that $\sim 2$. It should be noted
that for the $30 \%$ noise we chose, the $H(z)$ of
pure GR with a vanishing cosmological constant could have been
obtained. In that case, a potential-free scalar-tensor model with
$\Phi = {\rm const.}$ would of course have fitted perfectly this
$H(z)$ up to $z\rightarrow \infty$. The reason why we never managed
to go beyond $z_{\rm max} \sim 2$ is that we considered {\it random
noise\/}, instead of such a precise {\it bias\/} of our assumed
function $H(z)$, Eq.~(\ref{H2GR}). We are aware that our deformed
functions of Fig.~4 do not reproduce a realistic experimental noise.
However, they illustrate in a well defined way that an inaccurate
determination of $H(z)$ over a wide redshift interval is actually
more constraining than a precise measurement over a small
redshift interval only.

\vskip 1pc
The conclusion of the present subsection is therefore that a
scalar-tensor theory without potential can accommodate a Hubble
diagram consistent with (\ref{H2GR}), but only on a small
redshift interval if $\Omega_{\Lambda,0}$ is significant. The
experimental determination of the luminosity distance $D_L(z)$,
either accurately for $z \lesssim 1$ or even with large
(tens of percents) uncertainties up to redshifts $z\sim 2$,
severely constrains this subclass of theories. Future observations
should thus be able to distinguish them from general relativity,
and to confirm or rule them out without any ambiguity.

It is worth noting that such future determinations of $D_L(z)$ would
{\it a priori\/} be much more constraining than solar-system
experiments and binary pulsars tests. Indeed, although the
precision of the latter is quite impressive (see e.g.
\cite{willbook,def1,def4,def7}), they anyway probe only the first
two derivatives of $\ln A(\varphi)$, Eqs.~(\ref{dF0})-(\ref{beta0}),
whereas cosmological observations should give access to the full
shape of this function.

Let us also recall that the constraints we found crucially depend
on the fact that the theory should contain only positive-energy
excitations to be consistent, and notably that the function $F$
should remain always strictly positive. We did not use any other
cosmological observation, but obviously, once the microscopic
Lagrangian of a scalar-tensor theory has been reconstructed using
$D_L(z)$, all its other cosmological predictions should also be
checked. For instance, a bound $F_{\rm nuc} > 0.86\, F_0$ is given in
Ref.~\cite{bp99} for the value of the function $F$ at nucleosynthesis
time.\footnote{See however Ref.~\cite{dp99}, in which extremely small
values of $F_{\rm nuc}/F_0 = A^2_0 / A^2_{\rm nuc}$ are shown to be
consistent with the observed abundances of light elements, provided
$d^2A(\varphi)/d\varphi^2$ is large enough, where $A(\varphi)$ is
the matter-scalar coupling function (\ref{A}).} If one assumes that
$F(z)$ is monotonic, the reconstructed function of Fig.~1 would not
be consistent with this nucleosynthesis bound beyond $z\approx 0.3$.
This would be even more constraining than the bound $z < 0.68$ we
obtained just from mathematical consistency requirements.
Alternatively, a reconstructed function $F(z)$ like the one of Fig.~1
would be consistent with the above nucleosynthesis bound only if it
were non-monotonic beyond $z\gtrsim 0.6$. Although this would not be
forbidden from a purely phenomenological point of view, this would be
anyway unnatural, and more difficult to justify theoretically.

\subsection{Massless scalar field and an (arbitrary) nonzero
cosmological constant}
To confirm the results of the previous subsection, let us now
consider the case of a massless scalar field together with a
cosmological constant whose value {\it differs\/} from the one
entering our assumed $H(z)$, Eq.~(\ref{H2GR})-(\ref{OmegaValues}).
The question that we wish to address is the following: Can {\it part
of\/} the observed $\Omega_{\Lambda,0}$ be due to the presence of a
massless scalar field?

To impose a cosmological constant in a scalar-tensor theory, one
would naively choose a constant potential $U(\Phi)$ in action
(\ref{S_JF}). However, as shown by Eq.~(\ref{V}), the corresponding
potential $V(\varphi)$ of the helicity-0 degree of freedom
$\varphi$ would not be constant in this case (because $F(\Phi)$ is
{\it a priori\/} varying), and its second derivative would
give generically a nonvanishing scalar mass. To avoid any scalar
self-interaction, and in particular to set its mass to $0$, one
needs in fact to impose $V(\varphi) = {\rm const.}$ in the
Einstein-frame action (\ref{S_EF}). This defines a consistent
cosmological ``constant'' in a massless scalar-tensor theory.
Note that the corresponding Jordan-frame potential $U(\Phi)$ is then
proportional to $F^2(\Phi)$, and therefore that it does not
correspond to the usual notion of cosmological {\it constant\/} in
action (\ref{S_JF}).

Since our assumed ``observed'' $H(z)$ involves a parameter denoted
$\Omega_{\Lambda,0}$, Eqs.~(\ref{H2GR})-(\ref{OmegaValues}), let
us introduce a different notation for the contribution due to the
constant potential $V$~:
\begin{equation}
\Omega_{V,0}\equiv {2 F_0 V\over 3 H_0^2}~.
\label{OmegaV}
\end{equation}
It is easily checked that for $\Omega_{V,0} = \Omega_{\Lambda,0}$,
the solution $A(\varphi) = 1$ (or $F(\Phi) = 1$) is recovered, {\it
i.e.}, a scalar field minimally coupled to gravity with a constant
potential acting like a cosmological constant. Indeed, in terms of
the function $f(x)$ defined in (\ref{smallf}), Eq.~(\ref{F}) reads
\begin{equation}
(\Omega_{\Lambda,0}+\Omega_{m,0}\, x^3) x f''(x)
+ {3\over 2}\Omega_{m,0}\, x^3 f'(x)
-4 \Omega_{\kappa,0}\, x f(x)
- 6 \Omega_{V,0}\, x f^2(x)
= 3 \Omega_{m,0}~.
\label{EqfVconst}
\end{equation}
Note that this is now a {\it non}-linear equation in $F$, contrary
to Eq.~(\ref{Eqf}) above for the case of a vanishing potential.
If $\Omega_{V,0} = \Omega_{\Lambda,0}$, one finds that $f(x) =
x^{-2}$ is an obvious solution, {\it i.e.}, $F(z) = F_0 = {\rm
const.}$ A constant scalar field $\Phi$ (or $\varphi$) then
satisfies Eqs.~(\ref{Phi})--(\ref{varphiprimeBis}).

If we now consider a scalar-tensor theory for which $\Omega_{V,0}$
differs from the ``observed'' $\Omega_{\Lambda,0} \approx 0.7$, we
find that like in the previous subsection, there exists a maximum
redshift $z_{\rm max}$ beyond which $F(z)$ becomes negative, and
therefore beyond which the theory loses its mathematical
consistency. Figure 5 displays this maximum redshift as a function
of $\Omega_{V,0}$. We plot this figure for the initial condition
$F'_0 = 0$ ({\it i.e.}, for a theory which is presently
indistinguishable from GR in the solar system), but as before, we
verified that the curve is almost identical if one takes the maximum
value of $|F'_0|$ consistent with the solar-system bound
(\ref{dF0}). We also assume $\Omega_{\kappa,0} = 0$ (spatially flat
universe) for this figure, as we know from the previous discussion
that a value even as large as $|\Omega_{\kappa,0}| \sim 0.2$ does
not change qualitatively the results.

For $\Omega_{V,0} = 0$, we recover the result $z_{\rm max} \approx
0.66$ derived above for a vanishing potential. When $\Omega_{V,0}
< 0$, $z_{\rm max}$ becomes even smaller. As expected, this is worse
than in the potential-free case. On the contrary, when
$\Omega_{V,0}$ is positive ({\it i.e.}, when it contributes
positively to part of the ``observed'' $\Omega_{\Lambda,0}$), the
maximum redshift $z_{\rm max}$ increases. This is just due to the
fact that our massless scalar field needs to mimic a smaller
fraction of the ``observed'' $\Omega_{\Lambda,0}$, so that the
theory can remain consistent over a wider redshift interval.
However, we find that $z_{\rm max}$ is still smaller than $1.5$ for
$\Omega_{V,0}\leq 0.6$, and a $H(z)$ of the form
(\ref{H2GR})-(\ref{OmegaValues}) observed up to $z\sim 2$ would thus
suffice to rule out the model. If such a $H(z)$ could be confirmed
up to $z\sim 5$, one would need $\Omega_{V,0} \geq 0.694$ for our
massless scalar-tensor theory to fit it! Even so, the theory would
anyway become pathological at slightly higher redshifts. In
conclusion, a massless scalar {\it cannot\/} account for a
significant part of the observed cosmological constant if $H(z)$ is
experimentally found to be of the form (\ref{H2GR}) over a wide
redshift interval.

Let us note finally that for $\Omega_{V,0} > \Omega_{\Lambda,0}$,
Eq.~(\ref{EqfVconst}) does admit strictly positive solutions for
$f$ (or $F$) up to arbitrarily large redshifts. This ensures that
the graviton energy is always positive. However, it is now the
scalar field which needs to carry negative energy. Indeed,
Eq.~(\ref{varphiprimeBis}) gives a negative value for
$\varphi'^2$, basically because of the presence of the large
negative number $-U$ in this equation. [In the $Z(\Phi)=1$
parametrization, $\Phi'^2$ is obviously also negative, because of
Eq.~(\ref{positiveEnergy}), or directly from Eq.~(\ref{PhiBis})
which also involves a $-U$ term.]

Therefore, there is only one possibility for a consistent massless
scalar-tensor theory to reproduce (\ref{H2GR})
over a wide redshift interval: It must involve a cosmological
constant, whose contribution $\Omega_{V,0}$ is equal to (or very
slightly smaller than) the parameter $\Omega_{\Lambda,0}$ entering
(\ref{H2GR}). In other words, the theory should be extremely close
to GR plus a cosmological constant, and the massless scalar field
must have a negligible contribution. This illustrates again the main
conclusion of our paper: The experimental determination of
the luminosity distance $D_L(z)$ over a wide redshift interval,
up to $z\sim 2$, will suffice to rule out (or confirm) the existence
of a massless scalar partner to the graviton.

\subsection{Reconstruction of the potential $U$ from a given $F$}
In the previous two subsections, the matter-scalar coupling function
$F(\Phi)$ (or $A(\varphi)$) was reconstructed from the assumed
knowledge of $H(z)$, for theories whose potential $U(\Phi)$ (or
$V(\varphi)$) had a given form. We now consider the inverse problem.
We still assume that future observations will provide a Hubble diagram
consistent with (\ref{H2GR})-(\ref{OmegaValues}), but we wish now to
reconstruct the scalar-field potential $U$ for given forms of the
coupling function $F$.

\subsubsection{Generic scalar-tensor theories}
We first consider a generic two-parameter family of scalar-tensor
theories, which has already been studied in great detail for
solar-system, binary-pulsar and gravity-wave experiments
\cite{def7,def9}, as well as for cosmology starting with the
matter-dominated era \cite{dn93} and even back to nucleosynthesis
\cite{dp99}. Its definition is simplified if we work in the
Einstein frame (\ref{2.4})-(\ref{S_EF}). The matter-scalar coupling
function is simply given by
\begin{equation}
\ln A(\varphi) = \alpha_0(\varphi-\varphi_0) + {1\over
2}\beta_0(\varphi-\varphi_0)^2~,
\label{genericA}
\end{equation}
in which the present value of the scalar field, $\varphi_0$, may be
chosen to vanish without loss of generality. Any analytical function
$\ln A(\varphi)$ may be expanded in such a way, but we here assume
that no higher power of $\varphi$ appears, {\it i.e.}, that $\ln
A(\varphi)$ is strictly parabolic: It depends only on the two
parameters $\alpha_0$ and $\beta_0$. The latter is a simplified
notation for $(d\alpha/d\varphi)_0$, and should not be confused with
the post-Newtonian parameter $\beta$ defined in (\ref{betaPPN}).
[Actually, this equation shows that $\beta \approx 1+{1\over
2}\alpha_0^2\beta_0$.] Solar-system experiments impose $|\alpha_0| <
1.4\times 10^{-2}$, Eq.~(\ref{dF0}), while binary pulsars give
$\beta_0 > -4.5$, Eq.~(\ref{beta0}), for this class of theories. We
first study these models for the case of a spatially flat
universe ($\Omega_{\kappa,0} = 0$).

As shown by Eq.~(\ref{KG}), a constant scalar field $\varphi =
\varphi_0$ may be a solution if $\alpha_0 = 0$ (so that
$\alpha(\varphi)\propto (\varphi-\varphi_0)$ vanishes too) and if
the potential $V(\varphi)$ is also constant. Our assumed $H(z)$,
Eqs.~(\ref{H2GR})-(\ref{OmegaValues}), can thus always be
reproduced if the parameter $\alpha_0$ vanishes identically, and the
reconstructed potential merely reduces to the constant $V =
{3\over 2} H_0^2 \Omega_{\Lambda,0}$. This corresponds simply
to GR plus a cosmological constant, and the massless scalar degree
of freedom $\varphi$ remains unexcited, frozen at an extremum of
the parabola (\ref{genericA}). Actually, Eq.~(\ref{KG}) shows that
this extremum corresponds to a stable situation only if it is a
minimum, {\it i.e.}, if $\beta_0 \geq 0$ in (\ref{genericA}). This
is consistent with the results of Refs.~\cite{dn93,w98}: If the
theory involves a cosmological constant whose value equals the
``observed'' one in Eqs.~(\ref{H2GR})-(\ref{OmegaValues}), a massless
scalar field is cosmologically attracted towards a minimum of the
coupling function $\ln A(\varphi)$, and the present value of its
slope, $\alpha_0$, is expected to be generically very small.

On the other hand, if $\alpha_0$ is {\it not\/} assumed to vanish,
say if its value is comparable to the solar-system bound
(\ref{dF0}), then our reconstruction of the potential $V(\varphi)$
from Eqs.~(\ref{F})--(\ref{5.5}) leads to serious difficulties.
Their nature depends on the magnitude of the curvature parameter
$\beta_0$ of parabola (\ref{genericA}).

String-inspired models \cite{dp94} suggest that $\beta_0$ may be as
large as 10, or even 40. With such large values (and assuming
non-vanishing $\alpha_0$), our numerical integrations of
Eqs.~(\ref{F})--(\ref{5.5}) give concave potentials $V(\varphi)$,
unbounded from below. This corresponds to unstable theories, and
thereby to extremely fine-tuned initial conditions: Changing
slightly the derivative of the scalar field, $d\varphi/dz$, at high
redshifts would {\it a priori\/} yield a totally different universe
at present. This result tells us that this kind of models
cannot be consistent over a wide redshift interval with the exact
form of $H(z)$ we chose in (\ref{H2GR}), unless the parameter
$\alpha_0$ is extremely small. Actually, this is just another
way to present the results of Refs.~\cite{dn93,w98}: Since they
predict that $\alpha_0$ should be almost vanishing at present,
assuming a significant non-zero value implies that the theory is
unnatural.

To obtain convex-shaped potentials $V(\varphi)$ ({\it i.e.}, stable
theories) while still assuming a non-vanishing $\alpha_0$, we
typically need values of $|\beta_0| \lesssim 4$. However, the
reconstructed potentials always exhibit sudden changes of their
slope. Basically, they reproduce a cosmological constant over a
finite interval around $\varphi_0$ ({\it i.e.}, around $z = 0$), and
become rapidly divergent beyond a critical value of the scalar field
(depending on $\beta_0$). Therefore, as in subsection VI.B above, we
find that such scalar-tensor models can reproduce (\ref{H2GR}) only
if they involve a cosmological constant, whose
energy contribution is close to the parameter $\Omega_{\Lambda,0}$
entering $H(z)$, and if the scalar field has a negligible enough
influence. In other words, such models would not explain the small
but nonzero value of the observed cosmological constant by a
``quintessence'' mechanism, and would not be more natural than
merely assuming the existence of $\Lambda$.

The above results are significantly changed if we take into account
the possible spatial curvature of the universe. Indeed, smoother
potentials $V(\varphi)$ are obtained for closed universes
($\Omega_{\kappa,0} < 0$), and the present value of the
cosmological constant thus becomes more ``natural''.

To illustrate this feature, let us consider the case of a minimally
coupled scalar field (as in \cite{St98}), corresponding to
$\alpha_0 = \beta_0 = 0$ in Eq.~(\ref{genericA}). For an open
universe ($\Omega_{\kappa,0} > 0$), we find from Eq.~(\ref{5.5}) that
the scalar field would need to carry negative energy to reproduce
(\ref{H2GR}). On the other hand, for a closed universe
($\Omega_{\kappa,0} < 0$), one can derive analytically the
parametric form of the potential $V(\varphi)$. It can be expressed
in terms of the hypergeometric function ${}_2{\cal F}_1(a,b;c;x)$
(solution of the differential equation $x(1-x){\cal
F}''+[c-(a+b+1)x]{\cal F}' - ab {\cal F} = 0$):
\begin{mathletters}
\begin{eqnarray}
V &=& \left({3\over 2}\,\Omega_{\Lambda,0}+\Omega_{\kappa,0}\,
x^2\right)H_0^2~,
\label{Vofx}\\
\varphi &=& \pm\,x\,\sqrt{-{\Omega_{\kappa,0}\over
\Omega_{\Lambda,0}}}~{}_2{\cal F}_1\left({1\over 3},{1\over
2};{4\over 3}; -{\Omega_{m,0}\over \Omega_{\Lambda,0}}\,x^3\right)~,
\label{varphiofx}
\end{eqnarray}
\label{parametricV}
\end{mathletters}
where as before $x \equiv 1+z$. If $|\Omega_{\kappa,0}|$ is very
small, we recover that $V(\varphi)$ exhibits a sudden change of
slope, as was obtained above in the flat case. This is illustrated
by the left panel of Fig.~6. On the contrary, if
$|\Omega_{\kappa,0}|$ is large enough, the same analytical
expression (\ref{parametricV}) gives nice regular potentials, like
the one displayed in the right panel of Fig.~6. This reconstructed
$V(\varphi)$, as well as those obtained numerically for weakly
varying $\ln A(\varphi)$, Eq.~(\ref{genericA}), are natural in the
sense that they can be approximated by the exponential of simple
polynomials in $\varphi$. In that case, the observed value of the
cosmological constant does not appear as a mere parameter introduced
by hand in the Lagrangian, but corresponds basically to the present
value of $2V(\varphi_0)$. It should be noted that a value as large as
$\Omega_{\kappa,0} = -0.1$ is not excluded by the latest Boomerang
data, though it would be problematic in the framework of the
inflationary paradigm.

In conclusion, the existence of non-singular solutions over a long
period of time is again the constraining input. A non-minimally
coupled scalar field is essentially incompatible with (\ref{H2GR})
over a wide redshift interval, unless the scalar field is frozen at
a minimum of $\ln A(\varphi)$ (consistently with \cite{dn93,w98}).
If future experiments provide a Hubble diagram in accordance with
(\ref{H2GR}) and also give a very small value for
$\Omega_{\kappa,0}$, it will be possible to conclude that
scalar-tensor theories (either non-minimally or minimally coupled)
{\it cannot\/} explain in a natural way the existence of a
cosmological constant. On the other hand, if the universe is closed
and $|\Omega_{\kappa,0}|$ large enough, a ``quintessence'' mechanism
in a scalar-tensor theory seems more natural than a mere
cosmological constant.

\subsubsection{Scaling solutions}

The above conclusions can be confirmed by starting from a given
$F(z)$ (or $A(z)$), rather than $F(\Phi)$ (or $A(\varphi)$).
We consider here ``scaling solutions'', {\it i.e.}, we assume that
these functions behave as some power of the scale factor $a$. One
may for instance write $F(z) = (a/a_0)^p = (1+z)^{-p}$, with $~p
\geq 0$. As before, our aim is to reconstruct a regular potential
$V(\varphi)$ from the knowledge of $H(z)$,
assumed to be of the form (\ref{H2GR})-(\ref{OmegaValues}).

The strongest constraint on this class of theories is imposed by the
solar-system bound (\ref{dF0}). Indeed, using the definition
(\ref{alpha}) for $\alpha(\varphi)$, one can also write it as
$\alpha(\varphi) = - F'/(2\varphi'F)$, and Eq.~(\ref{5.5})
evaluated at $z=0$ then yields the following second-order equation
for $p$~:
\begin{equation}
(1-\alpha_0^2) p^2 - (2 + 3\Omega_{m,0}) \alpha_0^2 p +
4\Omega_{\kappa,0}\alpha_0^2 = 0~.
\label{Eq.p}
\end{equation}
Note that this equation does not depend on the full form of
Eq.~(\ref{H2GR}), but only on its first derivative at $z = 0$, {\it
i.e.}, on the deceleration parameter $q_0 = (H'/H)_0 -1$. The
constraints on $p$ derived below are thus valid as soon as
$q_0$ is of order $\sim -{1\over 2}$, consistently with the
estimated value (6.2) for $\Omega_{m,0}$.

In the case of a spatially flat universe ($\Omega_{\kappa,0} = 0$),
Eq.~(\ref{Eq.p}) gives immediately $p = (2 + 3 \Omega_{m,0})
\alpha_0^2 / (1-\alpha_0^2) \approx 3\alpha_0^2$, so that the
solar-system bound (\ref{dF0}) imposes $p < 6\times 10^{-4}$.
Therefore, the scalar field needs to be almost minimally coupled. If
$p$ vanishes identically, we recover as before the trivial solution
of GR plus a cosmological constant, together with an unexcited
minimally-coupled scalar field. On the other hand, if $p$ does not
vanish, one finds that the scalar field needs to carry negative
energy beyond $z\approx 1.4$. Even without trying to reconstruct the
potential $V(\varphi)$, one can thus conclude that such scaling
solutions would be ruled out by the observation of a $H(z)$
of the form (\ref{H2GR}) up to $z\sim 2$. Paradoxically, this result
is valid even for an infinitesimal (but nonzero) value of $p$.
Indeed, there exists a discontinuity between the case of a strictly
constant $F$ and that of a scaling solution $F(z) = (1+z)^{-p}$. At
first order in $p$, and still assuming $\Omega_{\kappa,0} = 0$, one
can write Eq.~(\ref{5.5}) as
\begin{equation}
2(1+z)^2\varphi'^2 = p
- 3 p \left( \ln (1+z) - {1\over2} \right)
{\Omega_{m,0} (1+z)^3 \over
\Omega_{\Lambda,0} + \Omega_{m,0} (1+z)^3}
+ O(p^2)~.
\label{psmall}
\end{equation}
This equation confirms that $\varphi'^2 \rightarrow 0$ when
$p \rightarrow 0$, and therefore that the scalar field tends towards
a constant in this limit. However, it carries positive
energy ($\varphi'^2\geq 0$) only if
\begin{equation}
(1+z)^3 \left( \ln (1+z) - {5\over 6} \right)
\leq {\Omega_{\Lambda,0}\over 3\Omega_{m,0}}~.
\label{psmallPositiveE}
\end{equation}
Since the right-hand side is estimated to be $\lesssim 1$, the
maximum value of $z$ is obtained for $\ln (1+z) \approx 5/6$, so
that the large numerical factor coming from $(1+z)^3$ in the
left-hand side is compensated by the small term inside the second
parentheses. Working iteratively, this maximum redshift can be
better approximated by $z_{\rm max}\approx e^{5/6} - 1 +
(\Omega_{\Lambda,0}/ 3\Omega_{m,0})e^{-5/3} \approx 1.45$, and the
actual numerical resolution of equality (\ref{psmallPositiveE})
for the values (\ref{OmegaValues}) gives $z_{\rm max} = 1.429$.
Therefore, even if $p$ is vanishingly small, a scaling solution
$F(z) = (1+z)^{-p}$ cannot be consistent with (\ref{H2GR}) beyond
this maximum redshift. This illustrates once
more that the experimental determination of $H(z)$ up to $z\sim 2$
would be more constraining that solar-system experiments for this
class of theories, provided one takes into account the requirement
of positive energy. Let us underline that the above value for
$z_{\rm max}$ is valid for a monomial $F = (a/a_0)^p$ but not for
more complicated polynomial expressions. Indeed, as shown for
instance in Sec.~VI.B above, there do exist scalar-tensor theories
consistent with (\ref{H2GR}) up to arbitrarily large redshifts, and
they do not need to be {\it strictly\/} equivalent to GR plus a
cosmological constant (although they must be close enough to it).
Moreover, the above maximum redshift is a consequence of the {\it
exact\/} form for $H(z)$ we chose in Eq.~(\ref{H2GR}). A slightly
different function may of course allow a positive-energy scalar
field up to much higher redshifts. It suffices that the right-hand
side of Eq.~(\ref{varphiprime}) be strictly positive for $F\approx
{\rm const.}$, and the case of a closed universe discussed below
provides an example, since the contribution $-\Omega_{\kappa,0}$ is
then positive in Eq.~(\ref{varphiprime}).

The case of a spatially open universe ($\Omega_{\kappa,0} > 0$) is
forbidden by Eq.~(\ref{Eq.p}), unless $\Omega_{\kappa,0}$ is
smaller than $\sim {1\over 2}\alpha_0^2 < 10^{-4}$. Such a
situation would be indistinguishable from the spatially flat case.

In a spatially closed universe ($\Omega_{\kappa,0} < 0$),
$p$ is given by the positive root of the second-order equation
Eq.~(\ref{Eq.p}). Remembering the solar-system bound (\ref{dF0}),
one may consider the case $\alpha_0^2 \ll |\Omega_{\kappa,0}|$,
and one gets $p\approx 2|\alpha_0|\sqrt{-\Omega_{\kappa,0}}$. Even
if one considered values of $\Omega_{\kappa,0}$ as large as $-0.1$,
this would limit $p$ to $\sim 10^{-2}$. Therefore, in this case
again, solar-system constraints impose that the scalar field should
be almost minimally coupled, if one looks for such scaling
solutions. The difference with the spatially flat case is that
Eqs.~(\ref{F})--(\ref{varphiprimeBis}) can now be integrated for any
redshift $z$ (from future infinity, $z=-1$, to arbitrarily large
$z$). Since $F(z)$ needs to be almost constant, we recover
solution (\ref{parametricV}) for the potential $V(\varphi)$. As
in Sec.~VI.C.1 above, we can thus conclude that such models would be
consistent with (\ref{H2GR}) over a wide redshift
interval only if they are (almost) minimally coupled, and they would
provide a natural ``quintessence'' mechanism to explain the
presently observed cosmological constant only if the universe is
(marginally) closed.

Let us end this paragraph by a remark concerning scaling solutions,
for which the scalar-field energy density scales like a power of $a$.
As mentioned in the Introduction, they have attracted a lot of
attention recently. For a minimally coupled field, the possible
scaling behaviors and the corresponding potentials can be
classified \cite{LiSc98}.
As for a non-minimally coupled field, a subclass of
theories was considered in \cite{Am99}, for which
\begin{equation}
U(\Phi)= C F(\Phi)^M,
\label{UFM}
\end{equation}
where $C$ and $M$ are constants. Since, besides these two constants,
there is only one unknown function of $\Phi$, the knowledge of
$H(z)$ suffices to reconstruct the full microscopic Lagrangian from
Eqs.~(\ref{F})--(\ref{5.5}) above. However, the main conclusion of
Ref.~\cite{Am99} can be recovered from a simple argument, without
any numerical integration. Indeed, it was shown in this reference
that there exists a universal behavior of these theories, depending
on $M$ but not on the precise shape of $F(\Phi)$. As emphasized in
\cite{Am99}, this result was obtained in the strong coupling limit,
corresponding formally to $Z(\Phi)\rightarrow 0$ in action
(\ref{S_JF}). Taking into account the assumed relation (\ref{UFM}),
the class of theories under consideration is thus defined by
\begin{equation}
S={1\over 16\pi G_*} \int d^4x \sqrt{-g}
\Bigl(F(\Phi)~R
- 2C F(\Phi)^M \Bigr)
+ S_m[\psi_m; g_{\mu\nu}]\ .
\label{S_JFZ0}
\end{equation}
If we now introduce a new scalar variable $\Psi = F(\Phi)$, we
notice that $\Phi$ disappears totally from the action. No
physical result can thus depend on the precise form of $F(\Phi)$,
and we recover the conclusion of \cite{Am99}. The constant $C$ may
also be set to 1 by a change of length units, and this class of
theories is thus parametrized by the single real number $M$. Any
physical prediction must therefore depend only on $M$.

\section{Conclusion}
In this work we have investigated the constraints that arise from
the {\it experimental} knowledge of the luminosity distance in
function of the redshift up to $z\sim 2$, corresponding to
$H(z)$ given by (\ref{H2GR})-(\ref{OmegaValues}). In particular, our
universe is then presently accelerating and we have studied the
viability of subclasses of scalar-tensor theories of gravity. We
have shown that the subclass of models in which the scalar partner
$\Phi$ of the graviton has no potential at all, and which satisfy
the present-day existing constraints, are inevitably ruled out if an
expansion of the form of Eq.~(\ref{H2GR}) holds even for a redshift
interval as tiny as $z < 2$ (see the precise numbers in section
VI). We see that these theories become pathological in the form of a
vanishing $F$, already at such low redshifts for which $H(z)$ will
be experimentally accessible in the near future \cite{SRSS}. Hence
we show that a cosmological observation of the background evolution
according to Eq.~(\ref{H2GR}) in the ``recent'' epoch will be enough
to rule out such models. [On the other hand, future observations
might provide a $H(z)$ which confirms the existence of a scalar
partner to the graviton and rule out pure GR!] The main reason why we
obtained so constraining results is that we took into account the
mathematical consistency of the theory, {\it i.e.}, the fact that it
should contain only positive-energy excitations to be well behaved.
This requirement severely restricts the class of viable models.

A non-flat universe can alleviate in some cases the tight
constraints we found. However, the latest CMB data released by
Boomerang and Maxima \cite{Bo,Ma} favor a flat universe (in
accordance with the inflationary paradigm), and only a marginally
closed universe is still allowed by the location of the first
acoustic (Doppler) peak at $\ell\sim 200$, while an open universe is
more unlikely.

The most impressive conclusion is that future cosmological
observations may prove to be more constraining for massless
scalar-tensor theories than solar-system and binary-pulsar tests.
Indeed, even if the determination of the luminosity distance
$D_L(z)$ will not reach very quickly the impressive accuracy
obtained in the solar system or with binary pulsars, it will
nevertheless give access to the full coupling function $F(\Phi)$ in
action (\ref{S_JF}), or $A(\varphi)$ in the Einstein-frame rewriting
(\ref{S_EF}), whereas only its first two derivatives are presently
probed.

\acknowledgments
We thank our collaborators A.~Starobinsky and B.~Boisseau for
discussions. One of us (GEF) also wishes to thank A.~Riazuelo for
informative discussions. Centre de Physique Th\'eorique is Unit\'e
Propre de Recherche 7061.

\begin{figure}
\begin{center}\leavevmode\epsfbox{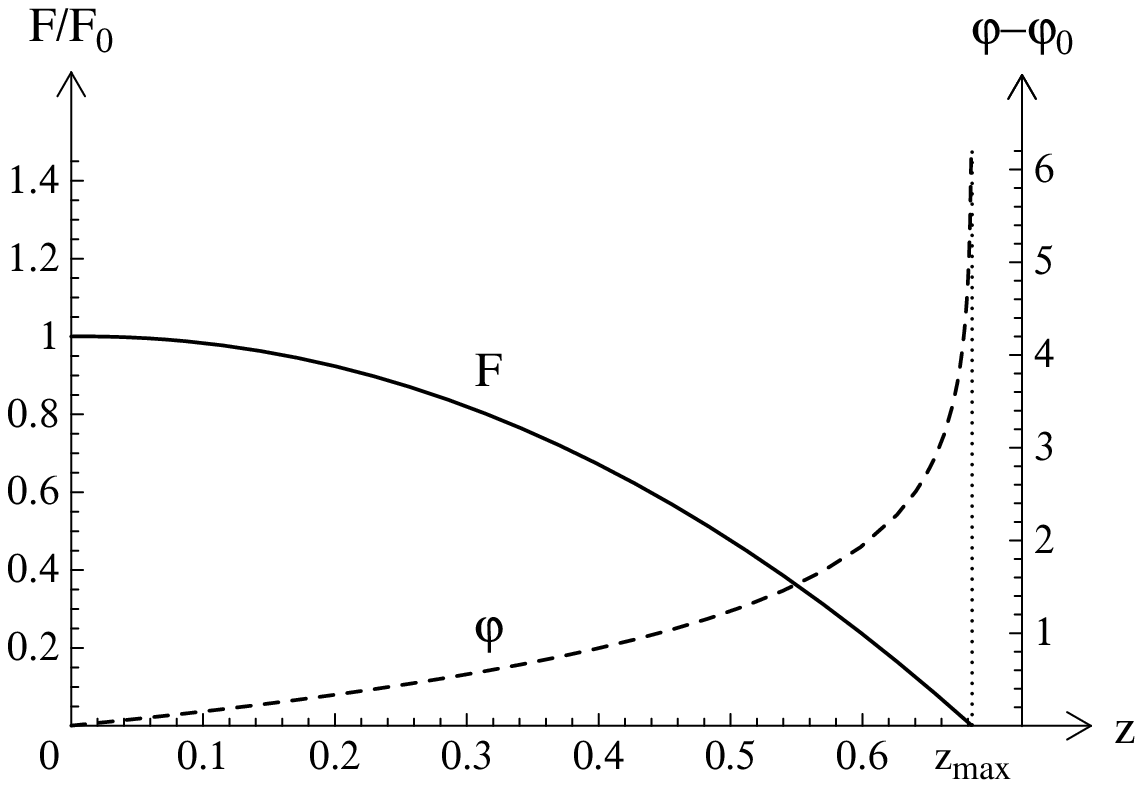}\end{center}\vskip 1pc
\caption{Reconstructed $F(z)$ [{\it i.e.}, Brans-Dicke
scalar $\Phi_{\rm BD}(z)$] and Einstein-frame scalar $\varphi$ as
functions of the Jordan-frame ({\it i.e.}, observed) redshift $z$, for
the maximum value of $|\alpha_0|$ allowed by solar-system experiments,
and for a vanishing potential. The helicity-0 degree of freedom
$\varphi$ diverges at $z_{\rm max} \approx 0.68$.}
\label{fig1}
\end{figure}

\begin{figure}
\begin{center}\leavevmode\epsfbox{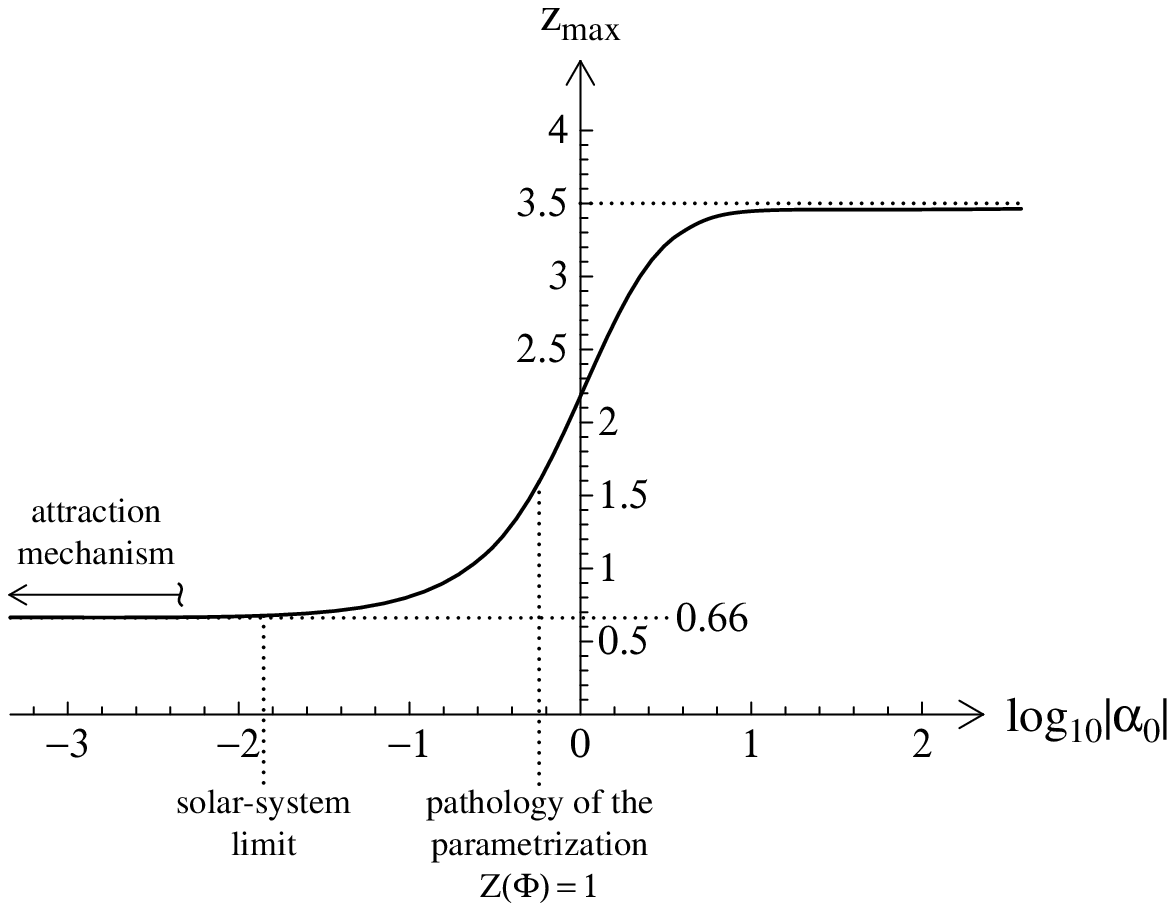}\end{center}\vskip 1pc
\caption{Maximum redshift $z$ consistent with the positivity of
energy of both the graviton and the scalar field, as a function of
the parameter $|\alpha_0|$. This figure corresponds to the case of a
vanishing scalar-field potential, and we fit the exact $H(z)$
predicted by general relativity plus a cosmological constant
(GR $+\Lambda$).}
\label{fig2}
\end{figure}

\begin{figure}
\begin{center}\leavevmode\epsfbox{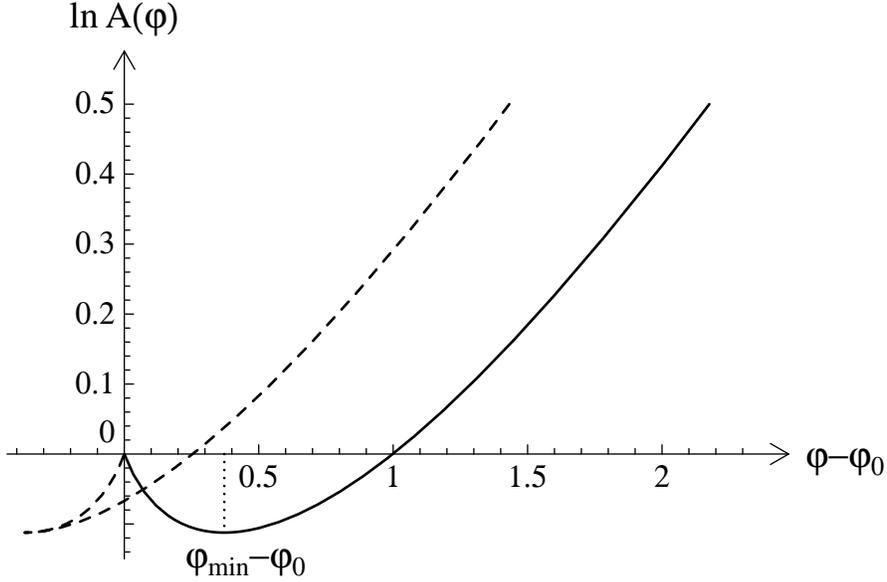}\end{center}\vskip 1pc
\caption{Two versions of the reconstructed coupling function $\ln
A(\varphi)$ for $|\alpha_0| = 1$, the dashed one looking bi-valued,
but the (single-valued) solid one giving the same predicted $H(z)$.
This figure still corresponds to the case of a vanishing scalar-field
potential, and we fit the exact $H(z)$ predicted by GR $+\Lambda$.}
\label{fig3}
\end{figure}

\begin{figure}
\begin{center}\leavevmode\epsfbox{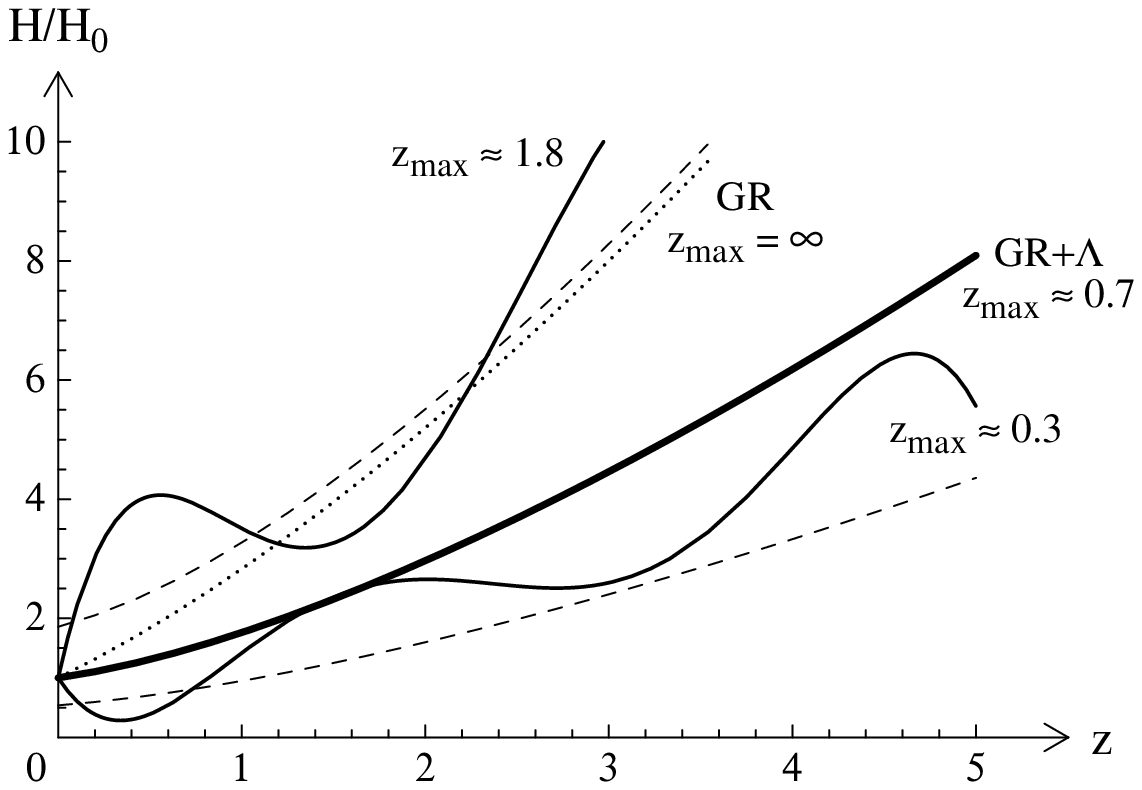}\end{center}\vskip 1pc
\caption{Random deformations of the $H(z)$ predicted by GR
$+\Lambda$ (with $\Omega_{\Lambda,0} = 0.7$), and corresponding
maximum value of the redshift $z$ consistent with the positivity of
energy. The dashed lines indicate the region in which random points
have been chosen at regular intervals of $z$. The thin solid lines
correspond to two polynomial fits of such random points. Note that
they can differ from the GR $+\Lambda$ curve even more than the
dashed lines. The dotted line labeled simply ``GR'' corresponds to a
vanishing cosmological constant $\Lambda$. Such a {\it bias\/} of
the GR $+\Lambda$ curve changes $z_{\rm max}$ much more that the
{\it random noise\/} we considered.}
\label{fig4}
\end{figure}

\begin{figure}
\begin{center}\leavevmode\epsfbox{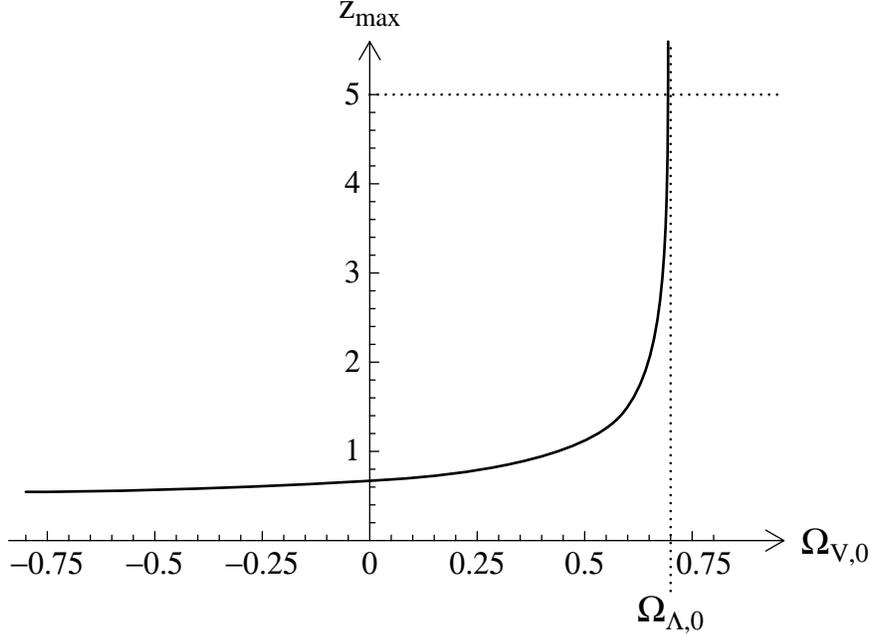}\end{center}\vskip 1pc
\caption{Maximum redshift $z$ consistent with the positivity of
energy, as a function of the value of a constant potential $V$ (case
of a massless helicity-0 degree of freedom $\varphi$).}
\label{fig5}
\end{figure}

\begin{figure}
\begin{center}\leavevmode\epsfxsize=\textwidth
\epsfbox{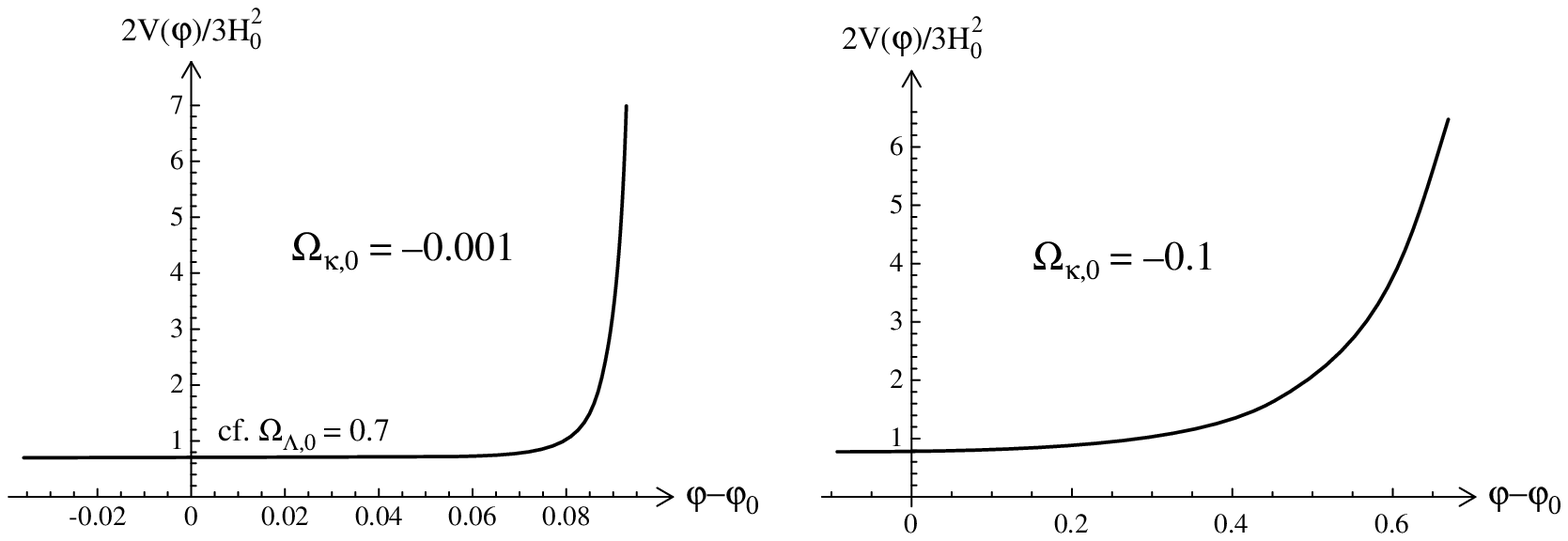}\end{center}\vskip 1pc
\caption{Minimally coupled model $F = 1$ in a spatially closed FRW
universe, respectively for $\Omega_{\kappa,0} = -10^{-3}$ (left
panel) and $\Omega_{\kappa,0} = -0.1$ (right panel). In both cases,
the potential $V(\varphi)$ is analytically given by
Eq.~(\protect{\ref{parametricV}}). Note that the reconstructed
potential does not have a ``natural'' shape if $|\Omega_{\kappa,0}|$
is too small: The present value of $\Omega_{\Lambda,0}$ is not
explained by a quintessence mechanism, and the corresponding
scalar-tensor theory is basically equivalent to GR $+\Lambda$. On
the contrary, if $|\Omega_{\Lambda,0}|$ is large enough, the
potential has a nice smooth shape, and its present value (at
$\varphi - \varphi_0 = 0$ on the figure) basically corresponds to
the observed $\Omega_{\Lambda,0}$.}
\label{fig6}
\end{figure}

\end{document}